\documentclass[%
reprint,
 showpacs,
 amsmath,amssymb,
]{revtex4-1}
\usepackage{graphicx}
\usepackage{dcolumn}
\usepackage{bm}

\usepackage{CJK}

\usepackage{color}
\usepackage{amssymb}
\usepackage{amsfonts}

\definecolor{Red}{rgb}{1,0,0}

\usepackage[colorlinks,urlcolor=blue,linkcolor=blue,citecolor=blue,anchorcolor=blue]{hyperref}
\makeatletter

\newcommand{\Rmnum}[1]{\expandafter\@slowromancap\romannumeral #1@}
\makeatother
\begin{document}

\preprint{APS/123-QED}

\title{Temporal-order-driven asymmetric quantum interference and temporal coherence enhancement in spontaneous six-wave mixing}


\author{Da Zhang$^{1,2}$}
\email{zhang1556433@sxnu.edu.cn}
\author{Yu Zhang$^1$}

\affiliation{%
$^1$\mbox{School of Physics and Electronic Engineering, Shanxi Normal University, Taiyuan 030031, China} \\
$^2$\mbox{Key Laboratory of Magnetic Molecules and Magnetic Information Materials, Ministry of Education, Taiyuan 030031, China} \\
}%

\date{\today}

\begin{abstract}
\noindent
Narrow-band multiphoton entanglement sources serve as a core enabling resource for advanced quantum information technologies.
Recently, researchers have directly generated energy-time entangled triphoton $W$ states in a hot atomic medium via spontaneous six-wave mixing for the first time.
However, a rigorous theoretical framework for this process remains lacking to date, confining our understanding to a mere extension of the biphoton model.
Here, we analytically investigate the generation mechanism of energy-time entangled triphotons and their classically controllable optical properties in an electromagnetically induced transparency-assisted five-level cold atomic system.
Notably, triphoton generation follows strict temporal ordering, resulting in asymmetric quantum interference in triple coincidence counts--unreplicable and unexplainable by the inherently symmetric biphoton model.
These results establish a rigorous physical framework for spontaneous six-wave mixing-generated triphotons, clarify their distinctions from states produced via cascaded nonlinear models, and substantially advance their utility in quantum information protocols.
\end{abstract}

\maketitle

Entanglement is a core resource for quantum information technology, with its nonlocal correlations granting quantum communication \cite{duan.nat.414.106500.2001,gisin2007quantum}, computation \cite{artur.RMP..68.733.1996,PhysRevLett.131.150601}, and sensing \cite{PhysRevLett.104.103602,pdlett.optica.7.2017} distinct advantages over classical systems.
In optical quantum systems, the energy-time degree of freedom exhibits an inherent high-dimensional Hilbert space, compatibility with existing optical infrastructure, and robust integration capability--rendering it ideal for high-dimensional quantum networks \cite{lu2023frequency,xavier2025energy}.
The primary approach to generating energy-time entanglement involves combining broadband entanglement with discretization control.
For instance, in bulk quantum systems, broadband frequency-entangled biphotons are discretized into high-dimensional modes using cavities \cite{PhysRevLett.83.2556,PhysRevLett.91.163602,PhysRevA.68.015803}, etalons \cite{maltese2020generation,xie2015harnessing}, and domain engineering \cite{morrison2022frequency}.
In integrated systems, microring resonators utilize resonant mode screening and strong field confinement, acting as the key component for achieving high-dimensional energy-time entanglement via spontaneous four-wave mixing (SFWM) \cite{kues2017chip,sugiura2020broadband,Jaramillo-Villegas:17,lu2022bayesian,PhysRevApplied.19.064026,mahmudlu2023fully}.
Despite significant progress in preparing and applying high-dimensional energy-time entanglement, the fundamental properties of the resulting states originate from these biphotons; without additional filtering, efficient interaction with atomic repeaters is hindered.

As an alternative standard method for energy-time entanglement generation, SFWM in atomic ensembles enables precise control of absorption, dispersion, and nonlinearity via intrinsic properties such as photo-induced atomic coherence.
The resulting biphotons exhibit long coherence times \cite{du.optica.2.2014}, high spectral brightness \cite{Harris.PRL.94.183601.2005,Harris.PRL.97.113602.2006,2016Subnatural,lin.prl.134.043602.2025}, and tunable entanglement dimensionality \cite{zhang.pra.96.053849.2017}, with applications in 85\%-efficient single-photon storage \cite{Yunfei2019Efficient}, quantum key distribution \cite{Liuchang.ol.8.2013}, quantum teleportation \cite{zhao.nature.430.54.2004}, and linear optical quantum computing \cite{knill2001scheme}.
Biphoton generation here involves rich physics: their wavepackets are governed by the interplay of third-order nonlinear susceptibility and phase mismatch function \cite{wen.pra.74.023808.2006,wen.74.023809.2006,wen.PRA.76.013825.2007,wen.pra.75.033809.2007,wen.PRA.77.033816.2008}.
The former clarifies the biphoton generation mechanism, while the latter coherently modulates natural spectral width via electromagnetically induced transparency (EIT)--surpassing the atomic natural linewidth to achieve subnatural linewidths \cite{2016Subnatural,lin.prl.134.043602.2025} and offering a promising route to advancing state-of-the-art quantum technologies.
Recently, spontaneous six-wave mixing (SSWM) was observed in hot atomic ensembles--representing a breakthrough for the single-step generation of energy-time entangled triphoton $W$ states \cite{li2024direct,feng2025observation}.
However, to date, no rigorous theoretical framework exists to describe this process, limiting our understanding of its physical nature.
While Ref. \cite{kangkang.aqt.35.2020,wu.pra.112.013706.2025} investigated SSWM across distinct energy levels using the perturbation chain model, their framework remains incomplete.
This incompleteness leads to conclusions that energy-time entanglement of triphotons is merely a trivial extension of that for biphotons, failing to fully reveal the underlying physics of triphotons.
More critically, energy-time entangled triphotons can also be produced by cascaded nonlinear models \cite{Wen:10,shalm.np.9.1.2012}; yet the essential distinction between the states generated via these two pathways remains unclear.

In this work, we study triphoton generation via EIT-assisted SSWM in a five-level cold atomic system.
Using Wen's methodology \cite{wen.PRA.76.013825.2007,wen.pra.75.033809.2007,wen.PRA.77.033816.2008} for studying light-atom interactions, we extract intrinsic frequency parameters--which govern the optical properties of triphoton wavepackets--from the linear and nonlinear susceptibilities of the generated signals.
The nonlinear susceptibility reveals that the entanglement dimension and structure of triphotons can be dynamically controlled by classical light.
Notably, triphoton generation exhibits strict temporal ordering, resulting in asymmetrically damped Rabi oscillations in triple coincidence counts--dominated by the fifth-order nonlinear susceptibility.
This phenomenon corresponds to genuine quantum interference, unexplainable by naturally symmetric biphotons.
In particular, such temporal ordering prolongs the coherence time of conditional biphotons.
By tuning characteristic frequency parameters, we shift triphoton wavepackets to a hybrid regime governed by the fifth-order nonlinear susceptibility and the longitudinal detuning function.
These results elucidate the triphoton generation mechanism and its dynamic coherent control process, while clarifying distinctions between SSWM-generated states and those from cascaded nonlinear models.

\begin{figure}[t]
\centering
  \includegraphics[width=8cm]{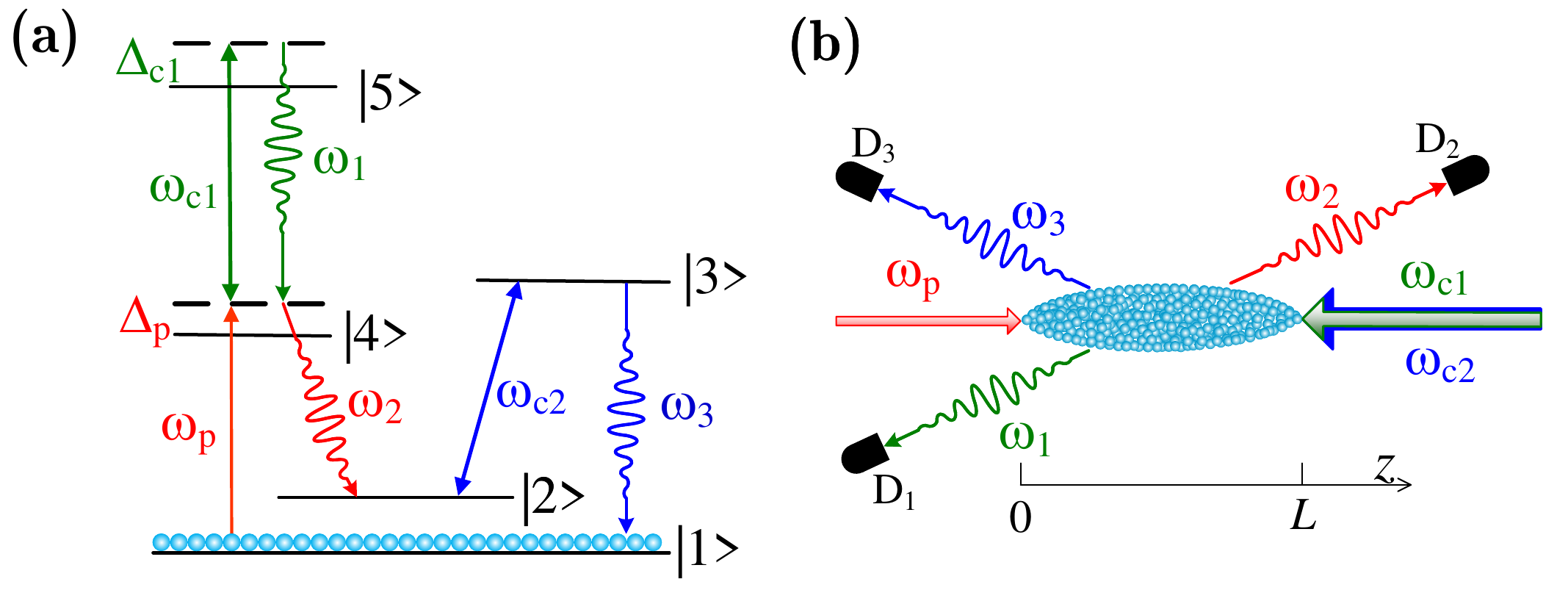}  
  \caption{(a) Triple-photon generation in a five-level cold atomic system. (b) Spatial configuration of incident light and generated signals. $D_i$ represents the $i$th detector.}
  \label{fig1}
\end{figure}

Our analysis starts by examining triphoton generation via SSWM in a five-level cold atomic system, as shown in Fig. \ref{fig1}(a).
These identically cooled atoms, confined to a thin cylindrical volume of length $L$ with atomic density $N$, are initially prepared in the ground state $|1\rangle$ via optical pumping.
Figure \ref{fig1}(b) depicts the spatial configuration of the classical and quantized fields involved in SSWM: a pump laser $E_p$ (frequency $\omega_p$, wavenumber $k_p$) and two coupling lasers $E_{c1}$ ($\omega_{c1}$, $k_{c1}$) and $E_{c2}$ ($\omega_{c2}$, $k_{c2}$) counterpropagate through the medium.
A weak pump field $E_p$ drives the transition $|1\rangle\rightarrow|4\rangle$ with frequency detuning $\Delta_p=\omega_{41}-\omega_p$; a strong coupling field $E_{c1}$ induces the transition $|4\rangle\rightarrow|5\rangle$ with detuning $\Delta_{c1}=\omega_{54}-\omega_{c1}$; and a resonant strong coupling field $E_{c2}$ couples the atomic transition $|2\rangle\rightarrow|3\rangle$.
$\omega_{ij}$ is the transition frequency from level $|j\rangle$ to $|i\rangle$.
Through the fifth-order nonlinear susceptibility $\chi^{(5)}$, phase-matched triphotons $E_1$ ($\omega_{1}$, $k_{1}$), $E_2$ ($\omega_{2}$, $k_{2}$) and $E_3$ ($\omega_{3}$, $k_{3}$) are generated by SSWM, which satisfies both energy and momentum conservation.
Notably, the coupling field $E_{c2}$ and $E_{3}$ photons form a standard three-level EIT scheme.
In this setup, $E_{c2}$ not only mediates SSWM process but also establishes a slow-light transparency window for $E_3$ photons.

We adopt the methodology developed by Wen et al. \cite{wen.PRA.76.013825.2007,wen.pra.75.033809.2007,wen.PRA.77.033816.2008}--a framework widely recognized for its efficacy in predicting the properties of nonclassical states generated via light-atom interactions--to investigate the dynamic evolution of the five-level system depicted in Fig. \ref{fig1}(a).
Under the dipole approximation and rotating-wave approximation, the Heisenberg evolution equations for the three primary atomic operators of interest are given by
\begin{align}\label{eq1}
\dot{Q}_{31}&=\Upsilon_{31}Q_{31}-id_{13}E_3^{(-)}-id_{23}E^*_{c2}Q_{21}+id_{14}E_p^*Q_{34}e^{i\Delta t} ,       \nonumber \\
\dot{Q}_{42}&=\Upsilon_{42}Q_{42}-id_{14}E_p^*Q_{12}e^{i\Delta t}+id_{23}E^*_{c2}Q_{43}      \nonumber \\
 &-id_{54}(E_{c1}+E_1^{(+)}e^{-i\Delta t})Q_{52},                                            \nonumber \\
 \dot{Q}_{54}&=\Gamma_{54}Q_{54}+id_{41}E_pQ_{51}+id_{42}E_2^{(+)}Q_{52}.
\end{align}
Here, $Q_{ij}$ is the transformed atomic operator, $d_{ij}$ is the dipole matrix element divided by $\hbar$, $\Upsilon_{31} =-i\delta_3-\gamma_{31}$, $\Gamma_{54} =i\Delta_{c1}-\gamma_{54}$, and $\Upsilon_{42}=-i\delta_2-\gamma_{42}$, where $\gamma_{ij}$ denotes the dephasing rate.
$\delta_i=\omega_i-\varpi_{i}$ describes a small frequency shift between the real frequency of $E_i$ photons and the central frequency $\varpi_{i}$ ($\varpi_{3}=\omega_{31}, \varpi_{2}=\omega_{42}$, and $\varpi_{1}=\omega_{54}-\Delta_p-\Delta_{c1}$).
The remaining evolution equations are detailed in Supplemental Material.
After relevant mathematical manipulations, we derive the linear and fifth-order nonlinear susceptibilities of the generated signal:
\begin{subequations}
\begin{align}
&\chi^{(5)}=\frac{-iN\hbar d_{41}d_{32}d_{24}d_{13}|d_{54}|^2\mathbb{T}^*_{51}}{\varepsilon_0(\Gamma^*_{41}\Gamma^*_{51}+\left|\Omega_{c1}\right|^2)D(\delta_2,\delta_3)}, \\
&D(\delta_2,\delta_3)=(\mathbb{T}^*_{41}\mathbb{T}^*_{51}+\left|\Omega_{c1}\right|^2)(\Upsilon^*_{21}\Upsilon^*_{31} +\left|\Omega_{c2}\right|^2),              \nonumber \\
&\chi_1(\delta_2,\delta_3)=\frac{-iN\hbar|d_{45}|^2|\Omega_{p}|^2\Omega_{c1}|^2}{\varepsilon_0\mathbb{T}_{54}(\Gamma_{41}\Gamma_{51}+\left|\Omega_{c1}\right|^2)(\mathbb{T}_{41}\mathbb{T}_{51}+\left|\Omega_{c1}\right|^2)}, \\
&\chi_2(\delta_2)=\frac{i\hbar N|d_{24}|^2|\Omega_p|^2\Gamma_{51}\mathbb{R}_{31}(\Upsilon^*_{52}\Upsilon^*_{53}+\left|\Omega_{c2}\right|^2)}{\varepsilon_0
(\Gamma_{41}\Gamma_{51}+\left|\Omega_{c1}\right|^2)(\mathbb{R}_{21}\mathbb{R}_{31}+\left|\Omega_{c2}\right|^2)}  \nonumber \\
&\times\frac{1}{(\Upsilon^*_{53}\left|\Omega_{c1}\right|^2+\Upsilon^*_{42}(\Upsilon^*_{52}\Upsilon^*_{53}+\left|\Omega_{c2}\right|^2))},  \\
&\chi_3(\delta_3)=\frac{-iN\hbar|d_{13}|^2}{\varepsilon_0(\Upsilon^*_{31}+\left|\Omega_{c2}\right|^2/\Upsilon^*_{21})},
\end{align}\label{eq2}
\end{subequations}
where $\Gamma_{41}=i\Delta_p-\gamma_{41}$, $\Gamma_{51}=i(\Delta_p+\Delta_{c1})-\gamma_{51}$, $\Upsilon_{21}=-i\delta_3-\gamma_{21}$, $\mathbb{T}_{ij}=\Gamma_{ij}-i(\Delta_p+\delta_2+\delta_3)$, $\mathbb{R}_{ij}=\Upsilon_{ij}+i(\Delta_p+\delta_2+\delta_3)$, $\Upsilon_{52}=i(\Delta_{c1}-\delta_2)-\gamma_{52}$, and $\Upsilon_{53}=i(\Delta_{c1}-\delta_2)-\gamma_{53}$.

We next analyze the atomic dipole oscillation at $\varpi_1-\delta_2-\delta_3$.
While Eq. (\ref{eq2}a) shows that the two-photon resonance condition $\Delta_p+\Delta_{c1}=0$ enhances SSWM efficiency, this condition complicates subsequent derivations.
To clarify the mechanism of triphoton generation, we thus set $\Delta_{c1}=0$.
Although $D(\delta_2,\delta_3)$ involves two unknown variables, the exact roots of its real part are readily obtainable.
Along the $\delta_3$ direction, two resonances emerge at $\Omega_{e2}/2$ and $-\Omega_{e2}/2$, with linewidth $\gamma_{e2}$, where $\Omega_{e2}=\sqrt{4|\Omega_{c2}|^2-(\gamma_{31}-\gamma_{21})^2}$ and $\gamma_{e2}=(\gamma_{21}+\gamma_{31})/2$.
Along the $-\delta_2-\delta_3=\delta_1$ direction, two resonances emerge at $\Omega_{e1}/2$ and $-\Omega_{e1}/2$, with linewidth $\gamma_{e1}$, where $\Omega_{e1}=\sqrt{4|\Omega_{c1}|^2-(\gamma_{41}-\gamma_{51})^2}$ and $\gamma_{e1}=(\gamma_{41}+\gamma_{51})/2$.
This indicates four distinct SSWM channels:
\begin{figure}[t]
\centering
  \includegraphics[width=8cm]{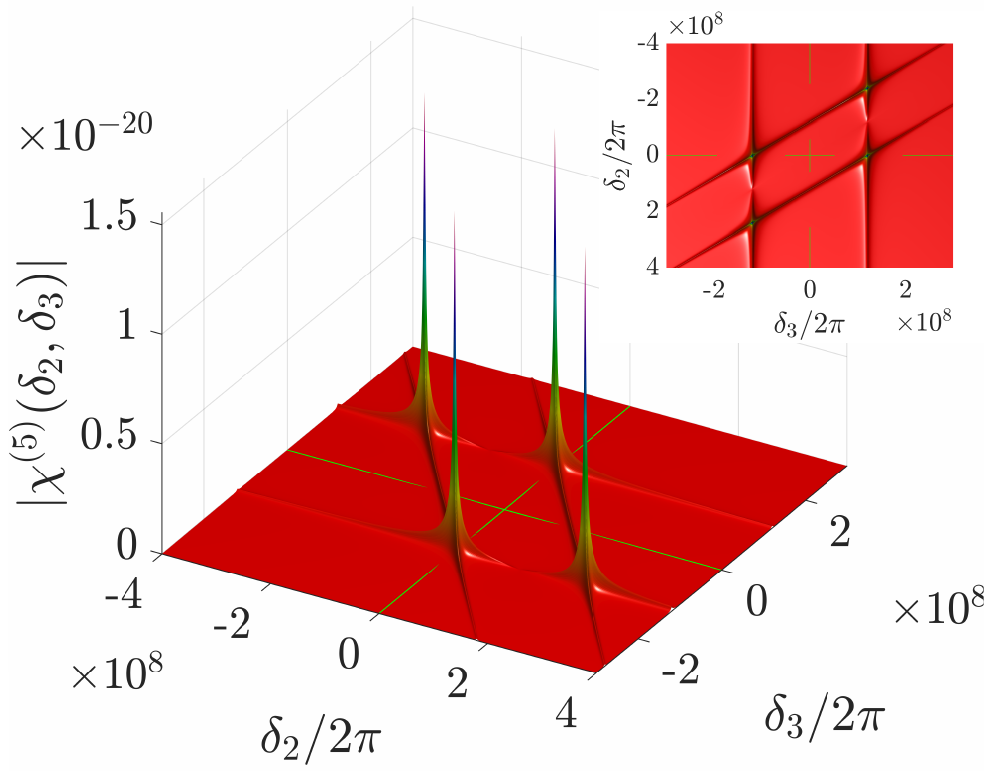}  
  \caption{The fifth-order nonlinear susceptibility $|\chi^{5}(\delta_2,\delta_3)|$ for $L=0.15$cm, $N=8.36\times10^{14}$, $\gamma_{31}=2\pi\times3$MHz, $\Omega_{c1}=\Omega_{c2}=40\gamma_{31}$, $\gamma_{41}=\gamma_{31}$, $\gamma_{51}=0.1\gamma_{31}$, $\gamma_{21}=0.02\gamma_{31}$, $\Delta_{c1}=\Delta_{c2}=0$, and $\Delta_p=-2\pi\times300$MHz.}
  \label{fig2}
\end{figure}
The first channel corresponds to peak frequencies $\varpi_1-\Omega_{e1}/2$ (for $\omega_1$), $\varpi_2+\Omega_{e1}/2+\Omega_{e2}/2$ (for $\omega_2$), and $\varpi_3-\Omega_{e2}/2$ (for $\omega_3$);
The second channel features peak frequencies $\varpi_1+\Omega_{e1}/2$, $\varpi_2-\Omega_{e1}/2+\Omega_{e2}/2$, and $\varpi_3-\Omega_{e2}/2$;
The third channel has peak frequencies $\varpi_1-\Omega_{e1}/2$, $\varpi_2+\Omega_{e1}/2-\Omega_{e2}/2$, and $\varpi_3+\Omega_{e2}/2$;
The fourth channel exhibits peak frequencies $\varpi_1+\Omega_{e1}/2$, $\varpi_2-\Omega_{e1}/2-\Omega_{e2}/2$, and $\varpi_3+\Omega_{e2}/2$.

Dimensionality of the generated state can be dynamically tuned based on $\Omega_{e1}$ and $\Omega_{e2}$.
For instance, when $\Omega_{e1}=\Omega_{e2}$, the state becomes a $2\times3\times2$-dimensional $W$ state in the frequency domain; tracing out any photon leaves the remaining two photons entangled.
When $\Omega_{e1}\neq\Omega_{e2}$, the generated state exhibits a $2\times4\times2$-dimensional structure but is not a $W$ state: tracing out $E_1$ or $E_3$ leaves the remaining two photons entangled, while tracing out $E_2$ renders the remaining bipartite subsystem disentangled.
This dynamic change in both entanglement dimensionality and structure highlights the notable tunability of our system.
Figure \ref{fig2} shows the spectral response of $|\chi^{(5)}|$ when $\Omega_{e1}=\Omega_{e2}$, where four resonant peaks are clearly resolved.
Notably, total spectral functions in cascaded nonlinear processes are always the product of two independent functions \cite{Wen:10,shalm.np.9.1.2012,hubel2010direct}, so their spectral responses consistently exhibit axial symmetry.
In contrast, energy conservation constraints (i.e., $\delta_1+\delta_2+\delta_3=0$) force $|\chi^{(5)}|$ to exhibit central symmetry around zero here.

In the interaction picture, the effective Hamiltonian for SSWM is given by
\begin{eqnarray}\label{eq3}
\hat{H}_I=\epsilon_0\int_Vd^3r\chi^{(5)}E_pE_{c1}E_{c2}E^{(-)}_{1}E^{(-)}_2E^{(-)}_3+H.c,
\end{eqnarray}
where \emph{V} denotes the interaction volume.
The three input fields are treated as classical, while the output signals are quantized.
For weak nonlinear interactions, the triphoton state output at the cell surface--derived by first-order perturbation theory--is
\begin{eqnarray}\label{eq4}
&&|\Psi\rangle = B_1\int d\delta_2d\delta_3\chi^{(5)}\Phi\delta(\Delta\omega)  |1,1,1\rangle,
\end{eqnarray}
where $B_1$ and $B_i$ (appearing in subsequent formulas) denoting constants.
Here, $\Phi=(1-e^{-i\Delta kL})/(i\Delta kL)$ is the longitudinal detuning function, where $\Delta k=k_p-k_{c1}-k_{c2}+k_1+k_{3}-k_{2}$.
The Dirac delta function $\delta(\Delta\omega)=\delta(\omega_p+\omega_{c1}+\omega_{c2}-\omega_1-\omega_2-\omega_3)$ ensures the generated triphotons exhibit energy-time tripartite entanglement for both discrete- and continuous-variable domains, while the entanglement structure and dimension from the discrete-variable perspective are given by $\chi^{(5)}$.

For $E_i$ in the atomic medium, the complex wavenumber is $k_i=(\omega_i/c)\sqrt{1+\chi_i}$, with its imaginary part accounting for Raman gain or loss.
For $|\Omega_p|^2\ll\Delta_p^2$, Eqs. (\ref{eq2}b)-(\ref{eq2}d) yield $k_{1}\simeq(\varpi_{1}-\delta_2-\delta_3)/c$, $k_{2}\simeq(\varpi_{1}+\delta_2)/c$, and $k_{3}=\varpi_{3}/c+\delta_3/\nu_{3}+i\varpi_3\mathrm{Im}[\chi_3]/c$ (the last term denotes EIT-induced loss).
The group velocity of $E_3$ is given by $\nu_3 =c/[1+\omega_{31}{\rm OD}\gamma_{31}/[2k_{31}L|\Omega_{c2}|^2]$, where OD$=N\zeta_{13}L$ is the optical depth and $\zeta_{13}=2\pi\hbar|d_{13}|^2/(\varepsilon_0\lambda_{13}\gamma_{31})$.
The wavenumber mismatch is thus expressed as
\begin{eqnarray}\label{eq5}
\Delta k = \frac{2(\omega_{21}-\Delta_p-\delta_2)}{c}+\delta_3(\frac{1}{\nu_{3}}+\frac{1}{c})+\frac{i\varpi_3\mathrm{Im}[\chi_3]}{c}.
\end{eqnarray}

We next consider the triple coincidence counts depicted in Fig. \ref{fig1}(b): $E_1$ (start photon) triggers detector $D_1$, with $E_2$ and $E_3$ (stop photons) triggering $D_2$ and $D_3$, respectively.
Because the triphoton bandwidth is much narrower than the spectral response bandwidth of single-photon detectors, the average triple coincidence rate is
\begin{eqnarray}\label{eq6}
&R_{cc}(\tau_{12},\tau_{13})=|\langle0|E^{(+)}_{3}(\tau_3)E^{(+)}_{2}(\tau_2)E^{(+)}_{1}(\tau_1)|\Psi\rangle|^2\nonumber \\
&=|\mathfrak{B}(\tau_{12},\tau_{13})|^2,
\end{eqnarray}
where $\tau_i=t_i-r_i/c$ with $r_i$ denoting the optical path from the medium's output surface to the detector $D_i$.
The function $\mathfrak{B}(\tau_{12},\tau_{13})$ is termed the triphoton wavepacket, where $\tau_{ij}=\tau_{j}-\tau_{i}$.
Substituting Eq. (\ref{eq4}) into Eq. (\ref{eq6}) yields
\begin{eqnarray}\label{eq7}
\mathfrak{B}(\tau_{12},\tau_{13})=B_2\int d\delta_2d\delta_3\chi^{(5)}\Phi e^{-i(\delta_2\tau_{12}+\delta_3\tau_{13})}.
\end{eqnarray}
It is the joint two-dimensional Fourier transform of $\chi^{(5)}$ and $\Phi$.
Four frequency parameters are extracted from $\chi^{(5)}$: $\Omega_{e1}$ and $\gamma_{e1}$ in the $\delta_2$ direction, and $\Omega_{e2}$ and $\gamma_{e2}$ in the $\delta_3$ direction.
Equation (\ref{eq5}) shows that $E_{c2}$ exclusively modulates the dispersion and transmission of $E_3$ in the $\delta_3$ direction--with the group delay-induced bandwidth $\Delta\omega_g \simeq 4\pi|\Omega_{c2}|^2/(\mathrm{OD}\gamma_{31})$ and transmission spectral width $\Delta\omega_t\simeq|\Omega_{c2}|^2/\sqrt{2\mathrm{OD}\gamma^2_{31}}$.
\begin{figure*}[htpb]
\centering
  \includegraphics[width=17.5cm]{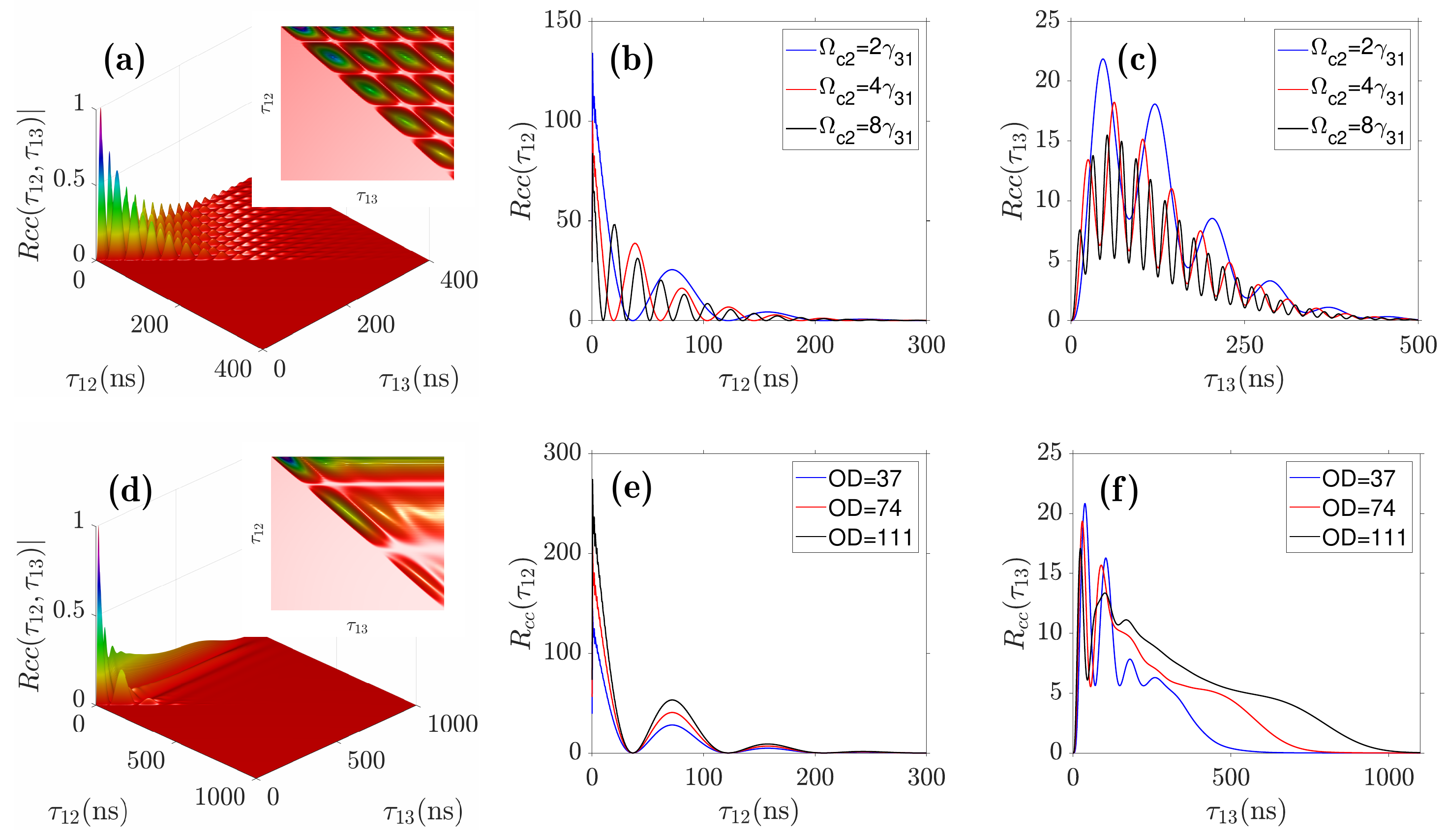}  
  \caption{(a) Normalized triple-coincidence counts $R_{cc}(\tau_{12},\tau_{13})$ dominated by $\chi^{(5)}$, with parameters as in Fig. \ref{fig2} except $\Omega_{c1}=\Omega_{c2}=8\gamma_{31}$. In this regime, (b) $R_{cc}(\tau_{12})$ and (c) $R_{cc}(\tau_{13})$ for $\Omega_{c1}=\Omega_{c2}=2\gamma_{31}$, $4\gamma_{31}$, and $8\gamma_{31}$, respectively.
  (d) Normalized $R_{cc}(\tau_{12},\tau_{13})$ dominated by both $\chi^{(5)}$ and $\Phi$, with parameters as in Fig. \ref{fig2} except $\Omega_{c1}=\Omega_{c2}=2\gamma_{31}$ and $OD=111$. (e) $R_{cc}(\tau_{12})$ and (f) $R_{cc}(\tau_{13})$ in the hybrid regime for OD=37, 74, 111, with remaining parameters same as in (d).}
  \label{fig3}
\end{figure*}
Thus, the competition between $2\gamma_{e2}$ and $\Delta\omega_g$ determines whether the profile of $\mathfrak{B}(\tau_{12},\tau_{13})$ in the $\tau_{13}$ direction is dominated by $\chi^{(5)}$ or $\Phi$.
In the $\delta_2$ direction, the natural spectral width $L/c\gg2\gamma_{e1}$, so its profile is dominated by $\chi^{(5)}$.

Analogously, the conditional two-photon coincidence rate is given by
\begin{eqnarray}\label{eq8}
R_{cc}(\tau_{1i})=B_3\int d\delta_j |\int d\delta_i\chi^{(5)}\Phi e^{-i(\delta_i\tau_{1i})}|^2.
\end{eqnarray}

We next consider a regime where the optical properties of $R_{cc}(\tau_{12},\tau_{13})$ are governed by $\chi^{(5)}$--a condition where $\Phi\approx1$ (requiring $2\gamma_{e2}<\Delta\omega_g$).
Substituting Eq. (\ref{eq2}a) into Eq. (\ref{eq7}) and after mathematical derivation, we obtain the triphoton wavepacket as
\begin{align}\label{eq9}
&\mathfrak{B}(\tau_{12},\tau_{13})=B_4e^{-\gamma_{e1}\tau_{12}-\gamma_{e2}(\tau_{13}-\tau_{12})}\Theta(\tau_{13}-\tau_{12})\Theta(\tau_{12})                                             \nonumber \\
&\times\left(\mathfrak{P}_{1}e^{i\frac{\Omega_{e1}\tau_{12}+\Omega_{e2}\left(\tau_{13}-\tau_{12}\right)}{2}}-\mathfrak{P}_{2}e^{i\frac{-\Omega_{e1}\tau_{12}+\Omega_{e2}\left(\tau_{13}-\tau_{12}\right)}{2}}-      \right.      \nonumber \\
&\left.\mathfrak{P}_{1}e^{i\frac{\Omega_{e1}\tau_{12}-\Omega_{e2}\left(\tau_{13}-\tau_{12}\right)}{2}}+\mathfrak{P}_{2}e^{i\frac{-\Omega_{e1}\tau_{12}-\Omega_{e2}\left(\tau_{13}-\tau_{12}\right)}{2}}\right),
\end{align}
where $\Theta(x)$ is the Heaviside step function, defined as $\Theta(x)=1$ for $x\geq0$ and $\Theta(x)=0$ for $x<0$.
Here, $\Theta(\tau_{12})$ means $E_1$ arrives at the detector not later than $E_2$, while $\Theta(\tau_{13}-\tau_{12})$ means $E_2$ arrives not later than $E_3$--indicating strict temporal ordering of triphoton generation, a physically intuitive feature.
Specifically, after the pump $E_p$ and coupling $E_{c1}$ excite atoms to level $|5\rangle$, $E_1$ is emitted first, followed by $E_2$; finally, after atoms are excited to $|3\rangle$, $E_3$ is generated.
Besides, the first to fourth terms on the right-hand side of Eq. (\ref{eq9}) correspond to the triphoton wavepackets of the four SSWM channels, with
$\mathfrak{P}_{1}=i(\Omega_{e1}/2-i\gamma_{e1})-\gamma_{51}$ and $\mathfrak{P}_{2}=i(-\Omega_{e1}/2-i\gamma_{e1})-\gamma_{51}$ denoting their respective weights.

The triple coincidence rate is expressed as:
\begin{align*}
&R_{cc}(\tau_{12},\tau_{13})=B_5e^{-2\gamma_{e1}\tau_{12}-2\gamma_{e2}(\tau_{13}-\tau_{12})}\left[\Omega_{e1}^2\cos^2\frac{\Omega_{e1}\tau_{12}}{2}   \right.
\end{align*}
\begin{align}\label{eq10}
&\left.+2\Omega_{e1}(\gamma_{51}-\gamma_{e1})\sin\Omega_{e1}\tau_{12} +4(\gamma_{51}-\gamma_{e1})^2\sin^2\frac{\Omega_{e1}\tau_{12}}{2} \right]      \nonumber \\
&\times[1-\cos\Omega_{e2}(\tau_{13}-\tau_{12})]\Theta^2(\tau_{13}-\tau_{12})\Theta^2(\tau_{12}).
\end{align}
Interference between four coherent SSWM channels induces damped Rabi oscillations in $R_{cc}(\tau_{12},\tau_{13})$, however, triphoton generation timing forces these oscillations to exhibit asymmetry in three-dimensional space--this constitutes genuine quantum interference unattainable with the biphoton model.
In contrast, for cascaded nonlinear processes, the corresponding triple coincidence rate always factorizes into $R_{cc}(\tau_{12})R_{cc}(\tau_{13})$ \cite{Wen:10,shalm.np.9.1.2012,hubel2010direct}, allowing reconstruction of the resulting triphoton wavepacket from its two marginal distributions (i.e., conditional two-photon coincidence rates).
In our system, by contrast, reconstruction of the triphoton wavepacket from such marginal distributions is impossible.

Figure \ref{fig3}(a) presents numerical results of normalized $R_{cc}(\tau_{12},\tau_{13})$ for the frequency-entangled triphoton $W$ state, using parameters as those in Fig. \ref{fig2} except $\Omega_{c1}=\Omega_{c2}=8\gamma_{31}$.
Calculations confirm $2\gamma_{e2}\ll\Delta\omega_g$, ensuring $R_{cc}(\tau_{12},\tau_{13})$ is governed by $\chi^{(5)}$.
Damped Rabi oscillations are resolved in Fig. \ref{fig3}(a), with a 21ns period in both $\tau_{12}$ and $\tau_{13}-\tau_{12}\geq0$ directions, and coherence times of 48ns [$1/(2\gamma_{e1})$] and 52ns [$1/(2\gamma_{e2})$], respectively.
Notably, numerical results agree excellently with Eq. (\ref{eq10}), validating our numerical model.

Additionally, the conditional two-photon coincidence rate is
\begin{align}\label{eq11}
R_{cc}(\tau_{12})&=\left[\Omega_{e1}\cos\frac{\Omega_{e1}\tau_{12}}{2}+2(\gamma_{51}-\gamma_{e1})\sin\frac{\Omega_{e1}\tau_{12}}{2}\right]^2 \nonumber  \\
&\times B_6e^{-2\gamma_{e1}\tau_{12}}\Theta(\tau_{12}).
\end{align}
Since the expression for $R_{cc}(\tau_{13})$ is lengthy, it is provided in Supplemental Material.

Figures \ref{fig3}(b) and \ref{fig3}(c) present numerical results for $R_{cc}(\tau_{12})$ and $R_{cc}(\tau_{13})$ with $\Omega_{c1}=\Omega_{c2}=2\gamma_{31}$, $4\gamma_{31}$, and $8\gamma_{31}$.
Since $E_3$ always lags behind $E_2$, the system after tracing out $E_3$ reduces to a SFWM system--quantum interference between two SFWM channels (with frequency $\omega_1=\varpi_1+\Omega_{e1}/2$, $\omega_2=\varpi_2-\Omega_{e1}/2$; $\omega_1=\varpi_1-\Omega_{e1}/2$, $\omega_2=\varpi_2+\Omega_{e1}/2$) induces damped Rabi oscillations in $R_{cc}(\tau_{12})$, with oscillation frequency $\Omega_{e1}$ and dephasing rate $2\gamma_{e1}$.
The behavior of $R_{cc}(\tau_{13})$ differs, however; tracing out $E_2$ effectively alters the characteristic frequency parameters along the $\tau_{13}$ direction, for instance, its damping rate becomes a superposition of $2\gamma_{e1}$ and $2\gamma_{e2}$.
As shown in Fig. \ref{fig3}(c), the coherence time of the conditional $E_1-E_3$ pairs is extended to 150ns, compared to 52ns for the triphoton case.
We refer to this phenomenon as \emph{temporal-order-driven coherence enhancement}.
Energy-time entanglement in these subsystems confirms the resulting state is indeed an energy-time entangled $W$ state.

We consider the hybrid regime, where $R_{cc}(\tau_{12},\tau_{13})$ is governed by $\chi^{(5)}$ along the $\tau_{12}$ direction and by $\Phi$ along the $\tau_{13}$ direction--this holds under $\Delta\omega_g<2\gamma_{e2}$.
Substituting Eqs. (\ref{eq5}) and (\ref{eq2}a) into Eq. (\ref{eq7}) yields
\begin{align}\label{eq12}
&\mathfrak{B}(\tau_{12},\tau_{13})= B_7(\frac{\Omega_{e1}}{2}\cos{\frac{\Omega_{e1}\tau_{12}}{2}}+\gamma_{e3}\sin{\frac{\Omega_{e1}\tau_{12}}{2}})\Theta(\tau_{12}) \nonumber \\
&\times\Theta(\tau_{13}-\tau_{12})\Pi(\tau_{13},L/\nu_3)e^{-\tau_{13}\mathbf{Im}[\chi_3]\varpi_3-\gamma_{e1}\tau_{12}},
\end{align}
where $\Pi(\tau_{13},L/\nu_3)$ is a rectangular function ranging from 0 to $L/\nu_3$.
In this regime, the frequency information of $E_3$ is erased by EIT filtering; thus, a rectangular of length $L/\nu_3$ appears along the $\tau_{13}$ direction for negligible EIT loss.
Besides, the frequency parameters along the $\tau_{12}$ direction remain unchanged--indicating the triphotons form a partially entangled state in the frequency domain.

Figure \ref{fig3}(d) shows $R_{cc}(\tau_{13},\tau_{23})$ in the hybrid regime, with parameters matching those in Fig. \ref{fig2} except OD=111 and $\Omega_{c1}=\Omega_{c2}=2\gamma_{31}$.
Calculations give $\Delta\omega_t\simeq$7.1MHz, and $\Delta\omega_g\simeq8.6$MHz$<2\gamma_{e2}$.
$R_{cc}(\tau_{13},\tau_{23})$ retains asymmetry owing to constraints from generation timing.
The sharp peak near $R_{cc}(0,0)$ corresponds to a triphoton-level precursor \cite{feng2025observation}, arising from triphotons outside the EIT window.
The tail along the $\tau_{13}$ direction exhibits exponential decay due to finite EIT loss, leading to a deviation from the ideal rectangular profile and thus prolonging the coherence time.
Figures \ref{fig3}(e) and \ref{fig3}(f) depict $R_{cc}(\tau_{12})$ and $R_{cc}(\tau_{13})$ for different OD, respectively.
The oscillation frequency and damping rate of $R_{cc}(\tau_{12})$ are invariant with OD, confirming the persistence of bipartite entanglement.
For $R_{cc}(\tau_{13})$, coherence times at OD=37, 74, and 111 are 245ns, 490ns, and 735ns, respectively.
Since the transmission spectrum--with a bandwidth narrower than $\Delta\omega_t$--further prolongs the coherence time of the conditional biphotons, this hybrid regime thus precludes time-ordered driven coherence enhancement.
Notably, the biphoton-level precursor remains resolvable and varies dynamically with OD \cite{du.prl.106.243602.2011,PhysRevLett.112.243602}.

In summary, we investigated energy-time entangled triphoton generation in a five-level cold atomic system.
The competition between $2\gamma_{e2}$ and $\Delta\omega_g$ dictates the optical properties of the triphoton wavepacket.
In the $2\gamma_{e2}$-dominated regime, four distinct SSWM channels enable dynamic control over the entanglement structure and dimensionality of triphotons via two coupling fields.
Notably, triphoton generation exhibits strict temporal ordering--resulting in asymmetric damped Rabi oscillations in triple coincidence counts, an effect unattainable with biphotons due to their inherent symmetry.
Specifically, this ordering extends the coherence time of conditional biphotons.
In the $\Delta\omega_g$-dominated hybrid regime, electromagnetically induced transparency erases discrete-variable energy-time tripartite entanglement, replacing it with a partially entangled tripartite state characterized by a subnatural linewidth.
Our contributions are threefold: 1) establishing a comprehensive theoretical framework for spontaneous six-wave mixing studies; 2) uncovering the underlying physics of triphoton generation and its dynamic coherent control mechanism; 3) clarifying distinctions between SSWM-generated triphotons and the states from cascaded nonlinear processes.
\section*{Acknowledgement}
This work was supported by the National Natural Science Foundation of China (12204293) and Applied Basic Research Program in Shanxi Province (No. 202203021212387).


\begin{thebibliography}{44}%
\makeatletter
\providecommand \@ifxundefined [1]{%
 \@ifx{#1\undefined}
}%
\providecommand \@ifnum [1]{%
 \ifnum #1\expandafter \@firstoftwo
 \else \expandafter \@secondoftwo
 \fi
}%
\providecommand \@ifx [1]{%
 \ifx #1\expandafter \@firstoftwo
 \else \expandafter \@secondoftwo
 \fi
}%
\providecommand \natexlab [1]{#1}%
\providecommand \enquote  [1]{``#1''}%
\providecommand \bibnamefont  [1]{#1}%
\providecommand \bibfnamefont [1]{#1}%
\providecommand \citenamefont [1]{#1}%
\providecommand \href@noop [0]{\@secondoftwo}%
\providecommand \href [0]{\begingroup \@sanitize@url \@href}%
\providecommand \@href[1]{\@@startlink{#1}\@@href}%
\providecommand \@@href[1]{\endgroup#1\@@endlink}%
\providecommand \@sanitize@url [0]{\catcode `\\12\catcode `\$12\catcode
  `\&12\catcode `\#12\catcode `\^12\catcode `\_12\catcode `\%12\relax}%
\providecommand \@@startlink[1]{}%
\providecommand \@@endlink[0]{}%
\providecommand \url  [0]{\begingroup\@sanitize@url \@url }%
\providecommand \@url [1]{\endgroup\@href {#1}{\urlprefix }}%
\providecommand \urlprefix  [0]{URL }%
\providecommand \Eprint [0]{\href }%
\providecommand \doibase [0]{http://dx.doi.org/}%
\providecommand \selectlanguage [0]{\@gobble}%
\providecommand \bibinfo  [0]{\@secondoftwo}%
\providecommand \bibfield  [0]{\@secondoftwo}%
\providecommand \translation [1]{[#1]}%
\providecommand \BibitemOpen [0]{}%
\providecommand \bibitemStop [0]{}%
\providecommand \bibitemNoStop [0]{.\EOS\space}%
\providecommand \EOS [0]{\spacefactor3000\relax}%
\providecommand \BibitemShut  [1]{\csname bibitem#1\endcsname}%
\let\auto@bib@innerbib\@empty
\bibitem [{\citenamefont {Duan}\ \emph {et~al.}(2001)\citenamefont {Duan},
  \citenamefont {Lukin}, \citenamefont {Cirac},\ and\ \citenamefont
  {Zoller}}]{duan.nat.414.106500.2001}%
  \BibitemOpen
  \bibfield  {author} {\bibinfo {author} {\bibfnamefont {L.}~\bibnamefont
  {Duan}}, \bibinfo {author} {\bibfnamefont {M.~D.}\ \bibnamefont {Lukin}},
  \bibinfo {author} {\bibfnamefont {J.~I.}\ \bibnamefont {Cirac}}, \ and\
  \bibinfo {author} {\bibfnamefont {P.}~\bibnamefont {Zoller}},\ }\href
  {http://www.nature.com/nature/journal/v414/n6862/full/414413a0.html}
  {\bibfield  {journal} {\bibinfo  {journal} {Nature}\ }\textbf {\bibinfo
  {volume} {414}},\ \bibinfo {pages} {413} (\bibinfo {year}
  {2001})}\BibitemShut {NoStop}%
\bibitem [{\citenamefont {Gisin}\ and\ \citenamefont
  {Thew}(2007)}]{gisin2007quantum}%
  \BibitemOpen
  \bibfield  {author} {\bibinfo {author} {\bibfnamefont {N.}~\bibnamefont
  {Gisin}}\ and\ \bibinfo {author} {\bibfnamefont {R.}~\bibnamefont {Thew}},\
  }\href {https://www.nature.com/articles/nphoton.2007.22} {\bibfield
  {journal} {\bibinfo  {journal} {Nat. Photon.}\ }\textbf {\bibinfo {volume}
  {1}},\ \bibinfo {pages} {165} (\bibinfo {year} {2007})}\BibitemShut {NoStop}%
\bibitem [{\citenamefont {Ekert}\ and\ \citenamefont
  {Jozsa}(1996)}]{artur.RMP..68.733.1996}%
  \BibitemOpen
  \bibfield  {author} {\bibinfo {author} {\bibfnamefont {A.}~\bibnamefont
  {Ekert}}\ and\ \bibinfo {author} {\bibfnamefont {R.}~\bibnamefont {Jozsa}},\
  }\href {\doibase 10.1103/RevModPhys.68.733} {\bibfield  {journal} {\bibinfo
  {journal} {Rev. Mod. Phys.}\ }\textbf {\bibinfo {volume} {68}},\ \bibinfo
  {pages} {733} (\bibinfo {year} {1996})}\BibitemShut {NoStop}%
\bibitem [{\citenamefont {Deng}\ \emph {et~al.}(2023)\citenamefont {Deng},
  \citenamefont {Gu}, \citenamefont {Liu}, \citenamefont {Gong}, \citenamefont
  {Su}, \citenamefont {Zhang}, \citenamefont {Tang}, \citenamefont {Jia},
  \citenamefont {Xu}, \citenamefont {Chen}, \citenamefont {Qin}, \citenamefont
  {Peng}, \citenamefont {Yan}, \citenamefont {Hu}, \citenamefont {Huang},
  \citenamefont {Li}, \citenamefont {Li}, \citenamefont {Chen}, \citenamefont
  {Jiang}, \citenamefont {Gan}, \citenamefont {Yang}, \citenamefont {You},
  \citenamefont {Li}, \citenamefont {Zhong}, \citenamefont {Wang},
  \citenamefont {Liu}, \citenamefont {Renema}, \citenamefont {Lu},\ and\
  \citenamefont {Pan}}]{PhysRevLett.131.150601}%
  \BibitemOpen
  \bibfield  {author} {\bibinfo {author} {\bibfnamefont {Y.-H.}\ \bibnamefont
  {Deng}}, \bibinfo {author} {\bibfnamefont {Y.-C.}\ \bibnamefont {Gu}},
  \bibinfo {author} {\bibfnamefont {H.-L.}\ \bibnamefont {Liu}}, \bibinfo
  {author} {\bibfnamefont {S.-Q.}\ \bibnamefont {Gong}}, \bibinfo {author}
  {\bibfnamefont {H.}~\bibnamefont {Su}}, \bibinfo {author} {\bibfnamefont
  {Z.-J.}\ \bibnamefont {Zhang}}, \bibinfo {author} {\bibfnamefont {H.-Y.}\
  \bibnamefont {Tang}}, \bibinfo {author} {\bibfnamefont {M.-H.}\ \bibnamefont
  {Jia}}, \bibinfo {author} {\bibfnamefont {J.-M.}\ \bibnamefont {Xu}},
  \bibinfo {author} {\bibfnamefont {M.-C.}\ \bibnamefont {Chen}}, \bibinfo
  {author} {\bibfnamefont {J.}~\bibnamefont {Qin}}, \bibinfo {author}
  {\bibfnamefont {L.-C.}\ \bibnamefont {Peng}}, \bibinfo {author}
  {\bibfnamefont {J.}~\bibnamefont {Yan}}, \bibinfo {author} {\bibfnamefont
  {Y.}~\bibnamefont {Hu}}, \bibinfo {author} {\bibfnamefont {J.}~\bibnamefont
  {Huang}}, \bibinfo {author} {\bibfnamefont {H.}~\bibnamefont {Li}}, \bibinfo
  {author} {\bibfnamefont {Y.}~\bibnamefont {Li}}, \bibinfo {author}
  {\bibfnamefont {Y.}~\bibnamefont {Chen}}, \bibinfo {author} {\bibfnamefont
  {X.}~\bibnamefont {Jiang}}, \bibinfo {author} {\bibfnamefont
  {L.}~\bibnamefont {Gan}}, \bibinfo {author} {\bibfnamefont {G.}~\bibnamefont
  {Yang}}, \bibinfo {author} {\bibfnamefont {L.}~\bibnamefont {You}}, \bibinfo
  {author} {\bibfnamefont {L.}~\bibnamefont {Li}}, \bibinfo {author}
  {\bibfnamefont {H.-S.}\ \bibnamefont {Zhong}}, \bibinfo {author}
  {\bibfnamefont {H.}~\bibnamefont {Wang}}, \bibinfo {author} {\bibfnamefont
  {N.-L.}\ \bibnamefont {Liu}}, \bibinfo {author} {\bibfnamefont {J.~J.}\
  \bibnamefont {Renema}}, \bibinfo {author} {\bibfnamefont {C.-Y.}\
  \bibnamefont {Lu}}, \ and\ \bibinfo {author} {\bibfnamefont {J.-W.}\
  \bibnamefont {Pan}},\ }\href {\doibase 10.1103/PhysRevLett.131.150601}
  {\bibfield  {journal} {\bibinfo  {journal} {Phys. Rev. Lett.}\ }\textbf
  {\bibinfo {volume} {131}},\ \bibinfo {pages} {150601} (\bibinfo {year}
  {2023})}\BibitemShut {NoStop}%
\bibitem [{\citenamefont {Anisimov}\ \emph {et~al.}(2010)\citenamefont
  {Anisimov}, \citenamefont {Raterman}, \citenamefont {Chiruvelli},
  \citenamefont {Plick}, \citenamefont {Huver}, \citenamefont {Lee},\ and\
  \citenamefont {Dowling}}]{PhysRevLett.104.103602}%
  \BibitemOpen
  \bibfield  {author} {\bibinfo {author} {\bibfnamefont {P.~M.}\ \bibnamefont
  {Anisimov}}, \bibinfo {author} {\bibfnamefont {G.~M.}\ \bibnamefont
  {Raterman}}, \bibinfo {author} {\bibfnamefont {A.}~\bibnamefont
  {Chiruvelli}}, \bibinfo {author} {\bibfnamefont {W.~N.}\ \bibnamefont
  {Plick}}, \bibinfo {author} {\bibfnamefont {S.~D.}\ \bibnamefont {Huver}},
  \bibinfo {author} {\bibfnamefont {H.}~\bibnamefont {Lee}}, \ and\ \bibinfo
  {author} {\bibfnamefont {J.~P.}\ \bibnamefont {Dowling}},\ }\href {\doibase
  10.1103/PhysRevLett.104.103602} {\bibfield  {journal} {\bibinfo  {journal}
  {Phys. Rev. Lett.}\ }\textbf {\bibinfo {volume} {104}},\ \bibinfo {pages}
  {103602} (\bibinfo {year} {2010})}\BibitemShut {NoStop}%
\bibitem [{\citenamefont {Anderson}\ \emph {et~al.}(2017)\citenamefont
  {Anderson}, \citenamefont {Gupta}, \citenamefont {Schmittberger},
  \citenamefont {Horrom}, \citenamefont {Hermann-Avigliano}, \citenamefont
  {Jones},\ and\ \citenamefont {Lett}}]{pdlett.optica.7.2017}%
  \BibitemOpen
  \bibfield  {author} {\bibinfo {author} {\bibfnamefont {B.~E.}\ \bibnamefont
  {Anderson}}, \bibinfo {author} {\bibfnamefont {P.}~\bibnamefont {Gupta}},
  \bibinfo {author} {\bibfnamefont {B.~L.}\ \bibnamefont {Schmittberger}},
  \bibinfo {author} {\bibfnamefont {T.}~\bibnamefont {Horrom}}, \bibinfo
  {author} {\bibfnamefont {C.}~\bibnamefont {Hermann-Avigliano}}, \bibinfo
  {author} {\bibfnamefont {K.~M.}\ \bibnamefont {Jones}}, \ and\ \bibinfo
  {author} {\bibfnamefont {P.~D.}\ \bibnamefont {Lett}},\ }\href
  {http://www.osapublishing.org/optica/abstract.cfm?URI=optica-4-7-752}
  {\bibfield  {journal} {\bibinfo  {journal} {Optica}\ }\textbf {\bibinfo
  {volume} {4}},\ \bibinfo {pages} {752} (\bibinfo {year} {2017})}\BibitemShut
  {NoStop}%
\bibitem [{\citenamefont {Lu}\ \emph {et~al.}(2023)\citenamefont {Lu},
  \citenamefont {Liscidini}, \citenamefont {Gaeta}, \citenamefont {Weiner},\
  and\ \citenamefont {Lukens}}]{lu2023frequency}%
  \BibitemOpen
  \bibfield  {author} {\bibinfo {author} {\bibfnamefont {H.-H.}\ \bibnamefont
  {Lu}}, \bibinfo {author} {\bibfnamefont {M.}~\bibnamefont {Liscidini}},
  \bibinfo {author} {\bibfnamefont {A.~L.}\ \bibnamefont {Gaeta}}, \bibinfo
  {author} {\bibfnamefont {A.~M.}\ \bibnamefont {Weiner}}, \ and\ \bibinfo
  {author} {\bibfnamefont {J.~M.}\ \bibnamefont {Lukens}},\ }\href
  {https://opg.optica.org/optica/fulltext.cfm?uri=optica-10-12-1655&id=544502}
  {\bibfield  {journal} {\bibinfo  {journal} {Optica}\ }\textbf {\bibinfo
  {volume} {10}},\ \bibinfo {pages} {1655} (\bibinfo {year}
  {2023})}\BibitemShut {NoStop}%
\bibitem [{\citenamefont {Xavier}\ \emph {et~al.}(2025)\citenamefont {Xavier},
  \citenamefont {Larsson}, \citenamefont {Villoresi}, \citenamefont {Vallone},\
  and\ \citenamefont {Cabello}}]{xavier2025energy}%
  \BibitemOpen
  \bibfield  {author} {\bibinfo {author} {\bibfnamefont {G.~B.}\ \bibnamefont
  {Xavier}}, \bibinfo {author} {\bibfnamefont {J.-{\AA}.}\ \bibnamefont
  {Larsson}}, \bibinfo {author} {\bibfnamefont {P.}~\bibnamefont {Villoresi}},
  \bibinfo {author} {\bibfnamefont {G.}~\bibnamefont {Vallone}}, \ and\
  \bibinfo {author} {\bibfnamefont {A.}~\bibnamefont {Cabello}},\ }\href
  {https://www.nature.com/articles/s41534-025-01072-3} {\bibfield  {journal}
  {\bibinfo  {journal} {NPJ Quant. Inform.}\ }\textbf {\bibinfo {volume}
  {11}},\ \bibinfo {pages} {129} (\bibinfo {year} {2025})}\BibitemShut
  {NoStop}%
\bibitem [{\citenamefont {Ou}\ and\ \citenamefont
  {Lu}(1999)}]{PhysRevLett.83.2556}%
  \BibitemOpen
  \bibfield  {author} {\bibinfo {author} {\bibfnamefont {Z.~Y.}\ \bibnamefont
  {Ou}}\ and\ \bibinfo {author} {\bibfnamefont {Y.~J.}\ \bibnamefont {Lu}},\
  }\href {\doibase 10.1103/PhysRevLett.83.2556} {\bibfield  {journal} {\bibinfo
   {journal} {Phys. Rev. Lett.}\ }\textbf {\bibinfo {volume} {83}},\ \bibinfo
  {pages} {2556} (\bibinfo {year} {1999})}\BibitemShut {NoStop}%
\bibitem [{\citenamefont {Lu}\ \emph {et~al.}(2003)\citenamefont {Lu},
  \citenamefont {Campbell},\ and\ \citenamefont {Ou}}]{PhysRevLett.91.163602}%
  \BibitemOpen
  \bibfield  {author} {\bibinfo {author} {\bibfnamefont {Y.~J.}\ \bibnamefont
  {Lu}}, \bibinfo {author} {\bibfnamefont {R.~L.}\ \bibnamefont {Campbell}}, \
  and\ \bibinfo {author} {\bibfnamefont {Z.~Y.}\ \bibnamefont {Ou}},\ }\href
  {\doibase 10.1103/PhysRevLett.91.163602} {\bibfield  {journal} {\bibinfo
  {journal} {Phys. Rev. Lett.}\ }\textbf {\bibinfo {volume} {91}},\ \bibinfo
  {pages} {163602} (\bibinfo {year} {2003})}\BibitemShut {NoStop}%
\bibitem [{\citenamefont {Goto}\ \emph {et~al.}(2003)\citenamefont {Goto},
  \citenamefont {Yanagihara}, \citenamefont {Wang}, \citenamefont {Horikiri},\
  and\ \citenamefont {Kobayashi}}]{PhysRevA.68.015803}%
  \BibitemOpen
  \bibfield  {author} {\bibinfo {author} {\bibfnamefont {H.}~\bibnamefont
  {Goto}}, \bibinfo {author} {\bibfnamefont {Y.}~\bibnamefont {Yanagihara}},
  \bibinfo {author} {\bibfnamefont {H.}~\bibnamefont {Wang}}, \bibinfo {author}
  {\bibfnamefont {T.}~\bibnamefont {Horikiri}}, \ and\ \bibinfo {author}
  {\bibfnamefont {T.}~\bibnamefont {Kobayashi}},\ }\href {\doibase
  10.1103/PhysRevA.68.015803} {\bibfield  {journal} {\bibinfo  {journal} {Phys.
  Rev. A}\ }\textbf {\bibinfo {volume} {68}},\ \bibinfo {pages} {015803}
  (\bibinfo {year} {2003})}\BibitemShut {NoStop}%
\bibitem [{\citenamefont {Maltese}\ \emph {et~al.}(2020)\citenamefont
  {Maltese}, \citenamefont {Amanti}, \citenamefont {Appas}, \citenamefont
  {Sinnl}, \citenamefont {Lemaitre}, \citenamefont {Milman}, \citenamefont
  {Baboux},\ and\ \citenamefont {Ducci}}]{maltese2020generation}%
  \BibitemOpen
  \bibfield  {author} {\bibinfo {author} {\bibfnamefont {G.}~\bibnamefont
  {Maltese}}, \bibinfo {author} {\bibfnamefont {M.}~\bibnamefont {Amanti}},
  \bibinfo {author} {\bibfnamefont {F.}~\bibnamefont {Appas}}, \bibinfo
  {author} {\bibfnamefont {G.}~\bibnamefont {Sinnl}}, \bibinfo {author}
  {\bibfnamefont {A.}~\bibnamefont {Lemaitre}}, \bibinfo {author}
  {\bibfnamefont {P.}~\bibnamefont {Milman}}, \bibinfo {author} {\bibfnamefont
  {F.}~\bibnamefont {Baboux}}, \ and\ \bibinfo {author} {\bibfnamefont
  {S.}~\bibnamefont {Ducci}},\ }\href {\doibase
  https://doi.org/10.1038/s41534-019-0237-9} {\enquote {\bibinfo {title}
  {Generation and symmetry control of quantum frequency combs},}\ } (\bibinfo
  {year} {2020})\BibitemShut {NoStop}%
\bibitem [{\citenamefont {Xie}\ \emph {et~al.}(2015)\citenamefont {Xie},
  \citenamefont {Zhong}, \citenamefont {Shrestha}, \citenamefont {Xu},
  \citenamefont {Liang}, \citenamefont {Gong}, \citenamefont {Bienfang},
  \citenamefont {Restelli}, \citenamefont {Shapiro}, \citenamefont {Wong} \emph
  {et~al.}}]{xie2015harnessing}%
  \BibitemOpen
  \bibfield  {author} {\bibinfo {author} {\bibfnamefont {Z.}~\bibnamefont
  {Xie}}, \bibinfo {author} {\bibfnamefont {T.}~\bibnamefont {Zhong}}, \bibinfo
  {author} {\bibfnamefont {S.}~\bibnamefont {Shrestha}}, \bibinfo {author}
  {\bibfnamefont {X.}~\bibnamefont {Xu}}, \bibinfo {author} {\bibfnamefont
  {J.}~\bibnamefont {Liang}}, \bibinfo {author} {\bibfnamefont {Y.-X.}\
  \bibnamefont {Gong}}, \bibinfo {author} {\bibfnamefont {J.~C.}\ \bibnamefont
  {Bienfang}}, \bibinfo {author} {\bibfnamefont {A.}~\bibnamefont {Restelli}},
  \bibinfo {author} {\bibfnamefont {J.~H.}\ \bibnamefont {Shapiro}}, \bibinfo
  {author} {\bibfnamefont {F.~N.}\ \bibnamefont {Wong}},  \emph {et~al.},\
  }\href {https://www.nature.com/articles/nphoton.2015.110} {\bibfield
  {journal} {\bibinfo  {journal} {Nat. Photon.}\ }\textbf {\bibinfo {volume}
  {9}},\ \bibinfo {pages} {536} (\bibinfo {year} {2015})}\BibitemShut {NoStop}%
\bibitem [{\citenamefont {Morrison}\ \emph {et~al.}(2022)\citenamefont
  {Morrison}, \citenamefont {Graffitti}, \citenamefont {Barrow}, \citenamefont
  {Pickston}, \citenamefont {Ho},\ and\ \citenamefont
  {Fedrizzi}}]{morrison2022frequency}%
  \BibitemOpen
  \bibfield  {author} {\bibinfo {author} {\bibfnamefont {C.~L.}\ \bibnamefont
  {Morrison}}, \bibinfo {author} {\bibfnamefont {F.}~\bibnamefont {Graffitti}},
  \bibinfo {author} {\bibfnamefont {P.}~\bibnamefont {Barrow}}, \bibinfo
  {author} {\bibfnamefont {A.}~\bibnamefont {Pickston}}, \bibinfo {author}
  {\bibfnamefont {J.}~\bibnamefont {Ho}}, \ and\ \bibinfo {author}
  {\bibfnamefont {A.}~\bibnamefont {Fedrizzi}},\ }\href
  {https://pubs.aip.org/aip/app/article/7/6/066102/2835141/Frequency-bin-entanglement-from-domain-engineered}
  {\bibfield  {journal} {\bibinfo  {journal} {APL Photon.}\ }\textbf {\bibinfo
  {volume} {7}} (\bibinfo {year} {2022})}\BibitemShut {NoStop}%
\bibitem [{\citenamefont {Kues}\ \emph {et~al.}(2017)\citenamefont {Kues},
  \citenamefont {Reimer}, \citenamefont {Roztocki}, \citenamefont {Cort{\'e}s},
  \citenamefont {Sciara}, \citenamefont {Wetzel}, \citenamefont {Zhang},
  \citenamefont {Cino}, \citenamefont {Chu}, \citenamefont {Little} \emph
  {et~al.}}]{kues2017chip}%
  \BibitemOpen
  \bibfield  {author} {\bibinfo {author} {\bibfnamefont {M.}~\bibnamefont
  {Kues}}, \bibinfo {author} {\bibfnamefont {C.}~\bibnamefont {Reimer}},
  \bibinfo {author} {\bibfnamefont {P.}~\bibnamefont {Roztocki}}, \bibinfo
  {author} {\bibfnamefont {L.~R.}\ \bibnamefont {Cort{\'e}s}}, \bibinfo
  {author} {\bibfnamefont {S.}~\bibnamefont {Sciara}}, \bibinfo {author}
  {\bibfnamefont {B.}~\bibnamefont {Wetzel}}, \bibinfo {author} {\bibfnamefont
  {Y.}~\bibnamefont {Zhang}}, \bibinfo {author} {\bibfnamefont
  {A.}~\bibnamefont {Cino}}, \bibinfo {author} {\bibfnamefont {S.~T.}\
  \bibnamefont {Chu}}, \bibinfo {author} {\bibfnamefont {B.~E.}\ \bibnamefont
  {Little}},  \emph {et~al.},\ }\href@noop {} {\bibfield  {journal} {\bibinfo
  {journal} {Nature}\ }\textbf {\bibinfo {volume} {546}},\ \bibinfo {pages}
  {622} (\bibinfo {year} {2017})}\BibitemShut {NoStop}%
\bibitem [{\citenamefont {Sugiura}\ \emph {et~al.}(2020)\citenamefont
  {Sugiura}, \citenamefont {Yin}, \citenamefont {Okamoto}, \citenamefont
  {Zhang}, \citenamefont {Kang}, \citenamefont {Chen}, \citenamefont {Wu},
  \citenamefont {Chu}, \citenamefont {Little},\ and\ \citenamefont
  {Takeuchi}}]{sugiura2020broadband}%
  \BibitemOpen
  \bibfield  {author} {\bibinfo {author} {\bibfnamefont {K.}~\bibnamefont
  {Sugiura}}, \bibinfo {author} {\bibfnamefont {Z.}~\bibnamefont {Yin}},
  \bibinfo {author} {\bibfnamefont {R.}~\bibnamefont {Okamoto}}, \bibinfo
  {author} {\bibfnamefont {L.}~\bibnamefont {Zhang}}, \bibinfo {author}
  {\bibfnamefont {L.}~\bibnamefont {Kang}}, \bibinfo {author} {\bibfnamefont
  {J.}~\bibnamefont {Chen}}, \bibinfo {author} {\bibfnamefont {P.}~\bibnamefont
  {Wu}}, \bibinfo {author} {\bibfnamefont {S.}~\bibnamefont {Chu}}, \bibinfo
  {author} {\bibfnamefont {B.}~\bibnamefont {Little}}, \ and\ \bibinfo {author}
  {\bibfnamefont {S.}~\bibnamefont {Takeuchi}},\ }\href
  {https://pubs.aip.org/aip/apl/article/116/22/224001/1022435/Broadband-generation-of-photon-pairs-from-a-CMOS}
  {\bibfield  {journal} {\bibinfo  {journal} {Appl. Phys. Lett.}\ }\textbf
  {\bibinfo {volume} {116}} (\bibinfo {year} {2020})}\BibitemShut {NoStop}%
\bibitem [{\citenamefont {Jaramillo-Villegas}\ \emph
  {et~al.}(2017)\citenamefont {Jaramillo-Villegas}, \citenamefont {Imany},
  \citenamefont {Odele}, \citenamefont {Leaird}, \citenamefont {Ou},
  \citenamefont {Qi},\ and\ \citenamefont {Weiner}}]{Jaramillo-Villegas:17}%
  \BibitemOpen
  \bibfield  {author} {\bibinfo {author} {\bibfnamefont {J.~A.}\ \bibnamefont
  {Jaramillo-Villegas}}, \bibinfo {author} {\bibfnamefont {P.}~\bibnamefont
  {Imany}}, \bibinfo {author} {\bibfnamefont {O.~D.}\ \bibnamefont {Odele}},
  \bibinfo {author} {\bibfnamefont {D.~E.}\ \bibnamefont {Leaird}}, \bibinfo
  {author} {\bibfnamefont {Z.-Y.}\ \bibnamefont {Ou}}, \bibinfo {author}
  {\bibfnamefont {M.}~\bibnamefont {Qi}}, \ and\ \bibinfo {author}
  {\bibfnamefont {A.~M.}\ \bibnamefont {Weiner}},\ }\href {\doibase
  10.1364/OPTICA.4.000655} {\bibfield  {journal} {\bibinfo  {journal} {Optica}\
  }\textbf {\bibinfo {volume} {4}},\ \bibinfo {pages} {655} (\bibinfo {year}
  {2017})}\BibitemShut {NoStop}%
\bibitem [{\citenamefont {Lu}\ \emph {et~al.}(2022)\citenamefont {Lu},
  \citenamefont {Myilswamy}, \citenamefont {Bennink}, \citenamefont {Seshadri},
  \citenamefont {Alshaykh}, \citenamefont {Liu}, \citenamefont {Kippenberg},
  \citenamefont {Leaird}, \citenamefont {Weiner},\ and\ \citenamefont
  {Lukens}}]{lu2022bayesian}%
  \BibitemOpen
  \bibfield  {author} {\bibinfo {author} {\bibfnamefont {H.-H.}\ \bibnamefont
  {Lu}}, \bibinfo {author} {\bibfnamefont {K.~V.}\ \bibnamefont {Myilswamy}},
  \bibinfo {author} {\bibfnamefont {R.~S.}\ \bibnamefont {Bennink}}, \bibinfo
  {author} {\bibfnamefont {S.}~\bibnamefont {Seshadri}}, \bibinfo {author}
  {\bibfnamefont {M.~S.}\ \bibnamefont {Alshaykh}}, \bibinfo {author}
  {\bibfnamefont {J.}~\bibnamefont {Liu}}, \bibinfo {author} {\bibfnamefont
  {T.~J.}\ \bibnamefont {Kippenberg}}, \bibinfo {author} {\bibfnamefont
  {D.~E.}\ \bibnamefont {Leaird}}, \bibinfo {author} {\bibfnamefont {A.~M.}\
  \bibnamefont {Weiner}}, \ and\ \bibinfo {author} {\bibfnamefont {J.~M.}\
  \bibnamefont {Lukens}},\ }\href
  {https://www.nature.com/articles/s41467-022-31639-z} {\bibfield  {journal}
  {\bibinfo  {journal} {Nat. Commun.}\ }\textbf {\bibinfo {volume} {13}},\
  \bibinfo {pages} {4338} (\bibinfo {year} {2022})}\BibitemShut {NoStop}%
\bibitem [{\citenamefont {Borghi}\ \emph {et~al.}(2023)\citenamefont {Borghi},
  \citenamefont {Tagliavacche}, \citenamefont {Sabattoli}, \citenamefont
  {Dirani}, \citenamefont {Youssef}, \citenamefont {Petit-Etienne},
  \citenamefont {Pargon}, \citenamefont {Sipe}, \citenamefont {Liscidini},
  \citenamefont {Sciancalepore}, \citenamefont {Galli},\ and\ \citenamefont
  {Bajoni}}]{PhysRevApplied.19.064026}%
  \BibitemOpen
  \bibfield  {author} {\bibinfo {author} {\bibfnamefont {M.}~\bibnamefont
  {Borghi}}, \bibinfo {author} {\bibfnamefont {N.}~\bibnamefont
  {Tagliavacche}}, \bibinfo {author} {\bibfnamefont {F.~A.}\ \bibnamefont
  {Sabattoli}}, \bibinfo {author} {\bibfnamefont {H.~E.}\ \bibnamefont
  {Dirani}}, \bibinfo {author} {\bibfnamefont {L.}~\bibnamefont {Youssef}},
  \bibinfo {author} {\bibfnamefont {C.}~\bibnamefont {Petit-Etienne}}, \bibinfo
  {author} {\bibfnamefont {E.}~\bibnamefont {Pargon}}, \bibinfo {author}
  {\bibfnamefont {J.}~\bibnamefont {Sipe}}, \bibinfo {author} {\bibfnamefont
  {M.}~\bibnamefont {Liscidini}}, \bibinfo {author} {\bibfnamefont
  {C.}~\bibnamefont {Sciancalepore}}, \bibinfo {author} {\bibfnamefont
  {M.}~\bibnamefont {Galli}}, \ and\ \bibinfo {author} {\bibfnamefont
  {D.}~\bibnamefont {Bajoni}},\ }\href {\doibase
  10.1103/PhysRevApplied.19.064026} {\bibfield  {journal} {\bibinfo  {journal}
  {Phys. Rev. Appl.}\ }\textbf {\bibinfo {volume} {19}},\ \bibinfo {pages}
  {064026} (\bibinfo {year} {2023})}\BibitemShut {NoStop}%
\bibitem [{\citenamefont {Mahmudlu}\ \emph {et~al.}(2023)\citenamefont
  {Mahmudlu}, \citenamefont {Johanning}, \citenamefont {Van~Rees},
  \citenamefont {Khodadad~Kashi}, \citenamefont {Epping}, \citenamefont
  {Haldar}, \citenamefont {Boller},\ and\ \citenamefont
  {Kues}}]{mahmudlu2023fully}%
  \BibitemOpen
  \bibfield  {author} {\bibinfo {author} {\bibfnamefont {H.}~\bibnamefont
  {Mahmudlu}}, \bibinfo {author} {\bibfnamefont {R.}~\bibnamefont {Johanning}},
  \bibinfo {author} {\bibfnamefont {A.}~\bibnamefont {Van~Rees}}, \bibinfo
  {author} {\bibfnamefont {A.}~\bibnamefont {Khodadad~Kashi}}, \bibinfo
  {author} {\bibfnamefont {J.~P.}\ \bibnamefont {Epping}}, \bibinfo {author}
  {\bibfnamefont {R.}~\bibnamefont {Haldar}}, \bibinfo {author} {\bibfnamefont
  {K.-J.}\ \bibnamefont {Boller}}, \ and\ \bibinfo {author} {\bibfnamefont
  {M.}~\bibnamefont {Kues}},\ }\href
  {https://www.nature.com/articles/s41566-023-01193-1} {\bibfield  {journal}
  {\bibinfo  {journal} {Nat. Photon.}\ }\textbf {\bibinfo {volume} {17}},\
  \bibinfo {pages} {518} (\bibinfo {year} {2023})}\BibitemShut {NoStop}%
\bibitem [{\citenamefont {Zhao}\ \emph {et~al.}(2014)\citenamefont {Zhao},
  \citenamefont {Guo}, \citenamefont {Liu}, \citenamefont {Sun}, \citenamefont
  {Loy},\ and\ \citenamefont {Du}}]{du.optica.2.2014}%
  \BibitemOpen
  \bibfield  {author} {\bibinfo {author} {\bibfnamefont {L.}~\bibnamefont
  {Zhao}}, \bibinfo {author} {\bibfnamefont {X.}~\bibnamefont {Guo}}, \bibinfo
  {author} {\bibfnamefont {C.}~\bibnamefont {Liu}}, \bibinfo {author}
  {\bibfnamefont {Y.}~\bibnamefont {Sun}}, \bibinfo {author} {\bibfnamefont
  {M.~M.~T.}\ \bibnamefont {Loy}}, \ and\ \bibinfo {author} {\bibfnamefont
  {S.}~\bibnamefont {Du}},\ }\href {\doibase 10.1364/OPTICA.1.000084}
  {\bibfield  {journal} {\bibinfo  {journal} {Optica}\ }\textbf {\bibinfo
  {volume} {1}},\ \bibinfo {pages} {84} (\bibinfo {year} {2014})}\BibitemShut
  {NoStop}%
\bibitem [{\citenamefont {Bali\ifmmode~\acute{c}\else \'{c}\fi{}}\ \emph
  {et~al.}(2005)\citenamefont {Bali\ifmmode~\acute{c}\else \'{c}\fi{}},
  \citenamefont {Braje}, \citenamefont {Kolchin}, \citenamefont {Yin},\ and\
  \citenamefont {Harris}}]{Harris.PRL.94.183601.2005}%
  \BibitemOpen
  \bibfield  {author} {\bibinfo {author} {\bibfnamefont {V.}~\bibnamefont
  {Bali\ifmmode~\acute{c}\else \'{c}\fi{}}}, \bibinfo {author} {\bibfnamefont
  {D.~A.}\ \bibnamefont {Braje}}, \bibinfo {author} {\bibfnamefont
  {P.}~\bibnamefont {Kolchin}}, \bibinfo {author} {\bibfnamefont {G.~Y.}\
  \bibnamefont {Yin}}, \ and\ \bibinfo {author} {\bibfnamefont {S.~E.}\
  \bibnamefont {Harris}},\ }\href {\doibase 10.1103/PhysRevLett.94.183601}
  {\bibfield  {journal} {\bibinfo  {journal} {Phys. Rev. Lett.}\ }\textbf
  {\bibinfo {volume} {94}},\ \bibinfo {pages} {183601} (\bibinfo {year}
  {2005})}\BibitemShut {NoStop}%
\bibitem [{\citenamefont {Kolchin}\ \emph {et~al.}(2006)\citenamefont
  {Kolchin}, \citenamefont {Du}, \citenamefont {Belthangady}, \citenamefont
  {Yin},\ and\ \citenamefont {Harris}}]{Harris.PRL.97.113602.2006}%
  \BibitemOpen
  \bibfield  {author} {\bibinfo {author} {\bibfnamefont {P.}~\bibnamefont
  {Kolchin}}, \bibinfo {author} {\bibfnamefont {S.}~\bibnamefont {Du}},
  \bibinfo {author} {\bibfnamefont {C.}~\bibnamefont {Belthangady}}, \bibinfo
  {author} {\bibfnamefont {G.~Y.}\ \bibnamefont {Yin}}, \ and\ \bibinfo
  {author} {\bibfnamefont {S.~E.}\ \bibnamefont {Harris}},\ }\href {\doibase
  10.1103/PhysRevLett.97.113602} {\bibfield  {journal} {\bibinfo  {journal}
  {Phys. Rev. Lett.}\ }\textbf {\bibinfo {volume} {97}},\ \bibinfo {pages}
  {113602} (\bibinfo {year} {2006})}\BibitemShut {NoStop}%
\bibitem [{\citenamefont {Shu}\ \emph {et~al.}(2016)\citenamefont {Shu},
  \citenamefont {Chen}, \citenamefont {Chow}, \citenamefont {Zhu},
  \citenamefont {Xiao}, \citenamefont {Loy},\ and\ \citenamefont
  {Du}}]{2016Subnatural}%
  \BibitemOpen
  \bibfield  {author} {\bibinfo {author} {\bibfnamefont {C.}~\bibnamefont
  {Shu}}, \bibinfo {author} {\bibfnamefont {P.}~\bibnamefont {Chen}}, \bibinfo
  {author} {\bibfnamefont {T.~K.~A.}\ \bibnamefont {Chow}}, \bibinfo {author}
  {\bibfnamefont {L.}~\bibnamefont {Zhu}}, \bibinfo {author} {\bibfnamefont
  {Y.}~\bibnamefont {Xiao}}, \bibinfo {author} {\bibfnamefont {M.~M.~T.}\
  \bibnamefont {Loy}}, \ and\ \bibinfo {author} {\bibfnamefont
  {S.}~\bibnamefont {Du}},\ }\href
  {https://www.nature.com/articles/ncomms12783} {\bibfield  {journal} {\bibinfo
   {journal} {Nat. Commun.}\ }\textbf {\bibinfo {volume} {7}},\ \bibinfo
  {pages} {12783} (\bibinfo {year} {2016})}\BibitemShut {NoStop}%
\bibitem [{\citenamefont {Lin}\ \emph {et~al.}(2025)\citenamefont {Lin},
  \citenamefont {Chien}, \citenamefont {Wu}, \citenamefont {Chinnarasu},
  \citenamefont {Du}, \citenamefont {Yu},\ and\ \citenamefont
  {Chuu}}]{lin.prl.134.043602.2025}%
  \BibitemOpen
  \bibfield  {author} {\bibinfo {author} {\bibfnamefont {J.-K.}\ \bibnamefont
  {Lin}}, \bibinfo {author} {\bibfnamefont {T.-H.}\ \bibnamefont {Chien}},
  \bibinfo {author} {\bibfnamefont {C.-T.}\ \bibnamefont {Wu}}, \bibinfo
  {author} {\bibfnamefont {R.}~\bibnamefont {Chinnarasu}}, \bibinfo {author}
  {\bibfnamefont {S.}~\bibnamefont {Du}}, \bibinfo {author} {\bibfnamefont
  {I.~A.}\ \bibnamefont {Yu}}, \ and\ \bibinfo {author} {\bibfnamefont {C.-S.}\
  \bibnamefont {Chuu}},\ }\href {\doibase 10.1103/PhysRevLett.134.043602}
  {\bibfield  {journal} {\bibinfo  {journal} {Phys. Rev. Lett.}\ }\textbf
  {\bibinfo {volume} {134}},\ \bibinfo {pages} {043602} (\bibinfo {year}
  {2025})}\BibitemShut {NoStop}%
\bibitem [{\citenamefont {Zhang}\ \emph {et~al.}(2017)\citenamefont {Zhang},
  \citenamefont {Zhang}, \citenamefont {Li}, \citenamefont {Zhang},
  \citenamefont {Cheng}, \citenamefont {Li},\ and\ \citenamefont
  {Zhang}}]{zhang.pra.96.053849.2017}%
  \BibitemOpen
  \bibfield  {author} {\bibinfo {author} {\bibfnamefont {D.}~\bibnamefont
  {Zhang}}, \bibinfo {author} {\bibfnamefont {Y.}~\bibnamefont {Zhang}},
  \bibinfo {author} {\bibfnamefont {X.}~\bibnamefont {Li}}, \bibinfo {author}
  {\bibfnamefont {D.}~\bibnamefont {Zhang}}, \bibinfo {author} {\bibfnamefont
  {L.}~\bibnamefont {Cheng}}, \bibinfo {author} {\bibfnamefont
  {C.}~\bibnamefont {Li}}, \ and\ \bibinfo {author} {\bibfnamefont
  {Y.}~\bibnamefont {Zhang}},\ }\href {\doibase 10.1103/PhysRevA.96.053849}
  {\bibfield  {journal} {\bibinfo  {journal} {Phys. Rev. A}\ }\textbf {\bibinfo
  {volume} {96}},\ \bibinfo {pages} {053849} (\bibinfo {year}
  {2017})}\BibitemShut {NoStop}%
\bibitem [{\citenamefont {Yunfei}\ \emph {et~al.}(2019)\citenamefont {Yunfei},
  \citenamefont {Wang}, \citenamefont {Jianfeng}, \citenamefont {Li},
  \citenamefont {Shanchao}, \citenamefont {Zhang}, \citenamefont {Keyu},
  \citenamefont {Su}, \citenamefont {Yiru},\ and\ \citenamefont
  {Zhou}}]{Yunfei2019Efficient}%
  \BibitemOpen
  \bibfield  {author} {\bibinfo {author} {\bibnamefont {Yunfei}}, \bibinfo
  {author} {\bibnamefont {Wang}}, \bibinfo {author} {\bibnamefont {Jianfeng}},
  \bibinfo {author} {\bibnamefont {Li}}, \bibinfo {author} {\bibnamefont
  {Shanchao}}, \bibinfo {author} {\bibnamefont {Zhang}}, \bibinfo {author}
  {\bibnamefont {Keyu}}, \bibinfo {author} {\bibnamefont {Su}}, \bibinfo
  {author} {\bibnamefont {Yiru}}, \ and\ \bibinfo {author} {\bibnamefont
  {Zhou}},\ }\href {https://www.nature.com/articles/s41566-019-0368-8}
  {\bibfield  {journal} {\bibinfo  {journal} {Nat. Photon.}\ }\textbf {\bibinfo
  {volume} {13}},\ \bibinfo {pages} {346} (\bibinfo {year} {2019})}\BibitemShut
  {NoStop}%
\bibitem [{\citenamefont {Liu}\ \emph {et~al.}(2013)\citenamefont {Liu},
  \citenamefont {Zhang}, \citenamefont {Zhao}, \citenamefont {Chen},
  \citenamefont {Fung}, \citenamefont {Chau}, \citenamefont {Loy},\ and\
  \citenamefont {Du}}]{Liuchang.ol.8.2013}%
  \BibitemOpen
  \bibfield  {author} {\bibinfo {author} {\bibfnamefont {C.}~\bibnamefont
  {Liu}}, \bibinfo {author} {\bibfnamefont {S.}~\bibnamefont {Zhang}}, \bibinfo
  {author} {\bibfnamefont {L.}~\bibnamefont {Zhao}}, \bibinfo {author}
  {\bibfnamefont {P.}~\bibnamefont {Chen}}, \bibinfo {author} {\bibfnamefont
  {C.~H.~F.}\ \bibnamefont {Fung}}, \bibinfo {author} {\bibfnamefont {H.~F.}\
  \bibnamefont {Chau}}, \bibinfo {author} {\bibfnamefont {M.~M.~T.}\
  \bibnamefont {Loy}}, \ and\ \bibinfo {author} {\bibfnamefont
  {S.}~\bibnamefont {Du}},\ }\href {\doibase 10.1364/OE.21.009505} {\bibfield
  {journal} {\bibinfo  {journal} {Opt. Express}\ }\textbf {\bibinfo {volume}
  {21}},\ \bibinfo {pages} {9505} (\bibinfo {year} {2013})}\BibitemShut
  {NoStop}%
\bibitem [{\citenamefont {Zhao}\ \emph {et~al.}(2004)\citenamefont {Zhao},
  \citenamefont {Chen}, \citenamefont {Zhang}, \citenamefont {Yang},
  \citenamefont {Briegel},\ and\ \citenamefont
  {Pan}}]{zhao.nature.430.54.2004}%
  \BibitemOpen
  \bibfield  {author} {\bibinfo {author} {\bibfnamefont {Z.}~\bibnamefont
  {Zhao}}, \bibinfo {author} {\bibfnamefont {Y.}~\bibnamefont {Chen}}, \bibinfo
  {author} {\bibfnamefont {A.}~\bibnamefont {Zhang}}, \bibinfo {author}
  {\bibfnamefont {T.}~\bibnamefont {Yang}}, \bibinfo {author} {\bibfnamefont
  {H.~J.}\ \bibnamefont {Briegel}}, \ and\ \bibinfo {author} {\bibfnamefont
  {J.}~\bibnamefont {Pan}},\ }\href
  {http://www.nature.com/nature/journal/v430/n6995/full/nature02643.html}
  {\bibfield  {journal} {\bibinfo  {journal} {Nature}\ }\textbf {\bibinfo
  {volume} {430}},\ \bibinfo {pages} {54} (\bibinfo {year} {2004})}\BibitemShut
  {NoStop}%
\bibitem [{\citenamefont {Knill}\ \emph {et~al.}(2001)\citenamefont {Knill},
  \citenamefont {Laflamme},\ and\ \citenamefont {Milburn}}]{knill2001scheme}%
  \BibitemOpen
  \bibfield  {author} {\bibinfo {author} {\bibfnamefont {E.}~\bibnamefont
  {Knill}}, \bibinfo {author} {\bibfnamefont {R.}~\bibnamefont {Laflamme}}, \
  and\ \bibinfo {author} {\bibfnamefont {G.~J.}\ \bibnamefont {Milburn}},\
  }\href {https://www.nature.com/articles/35051009} {\bibfield  {journal}
  {\bibinfo  {journal} {Nature}\ }\textbf {\bibinfo {volume} {409}},\ \bibinfo
  {pages} {46} (\bibinfo {year} {2001})}\BibitemShut {NoStop}%
\bibitem [{\citenamefont {Wen}\ and\ \citenamefont
  {Rubin}(2006{\natexlab{a}})}]{wen.pra.74.023808.2006}%
  \BibitemOpen
  \bibfield  {author} {\bibinfo {author} {\bibfnamefont {J.}~\bibnamefont
  {Wen}}\ and\ \bibinfo {author} {\bibfnamefont {M.~H.}\ \bibnamefont
  {Rubin}},\ }\href {\doibase 10.1103/PhysRevA.74.023808} {\bibfield  {journal}
  {\bibinfo  {journal} {Phys. Rev. A}\ }\textbf {\bibinfo {volume} {74}},\
  \bibinfo {pages} {023808} (\bibinfo {year} {2006}{\natexlab{a}})}\BibitemShut
  {NoStop}%
\bibitem [{\citenamefont {Wen}\ and\ \citenamefont
  {Rubin}(2006{\natexlab{b}})}]{wen.74.023809.2006}%
  \BibitemOpen
  \bibfield  {author} {\bibinfo {author} {\bibfnamefont {J.}~\bibnamefont
  {Wen}}\ and\ \bibinfo {author} {\bibfnamefont {M.~H.}\ \bibnamefont
  {Rubin}},\ }\href {\doibase 10.1103/PhysRevA.74.023809} {\bibfield  {journal}
  {\bibinfo  {journal} {Phys. Rev. A}\ }\textbf {\bibinfo {volume} {74}},\
  \bibinfo {pages} {023809} (\bibinfo {year} {2006}{\natexlab{b}})}\BibitemShut
  {NoStop}%
\bibitem [{\citenamefont {Wen}\ \emph {et~al.}(2007{\natexlab{a}})\citenamefont
  {Wen}, \citenamefont {Du},\ and\ \citenamefont
  {Rubin}}]{wen.PRA.76.013825.2007}%
  \BibitemOpen
  \bibfield  {author} {\bibinfo {author} {\bibfnamefont {J.}~\bibnamefont
  {Wen}}, \bibinfo {author} {\bibfnamefont {S.}~\bibnamefont {Du}}, \ and\
  \bibinfo {author} {\bibfnamefont {M.~H.}\ \bibnamefont {Rubin}},\ }\href
  {\doibase 10.1103/PhysRevA.76.013825} {\bibfield  {journal} {\bibinfo
  {journal} {Phys. Rev. A}\ }\textbf {\bibinfo {volume} {76}},\ \bibinfo
  {pages} {013825} (\bibinfo {year} {2007}{\natexlab{a}})}\BibitemShut
  {NoStop}%
\bibitem [{\citenamefont {Wen}\ \emph {et~al.}(2007{\natexlab{b}})\citenamefont
  {Wen}, \citenamefont {Du},\ and\ \citenamefont
  {Rubin}}]{wen.pra.75.033809.2007}%
  \BibitemOpen
  \bibfield  {author} {\bibinfo {author} {\bibfnamefont {J.}~\bibnamefont
  {Wen}}, \bibinfo {author} {\bibfnamefont {S.}~\bibnamefont {Du}}, \ and\
  \bibinfo {author} {\bibfnamefont {M.~H.}\ \bibnamefont {Rubin}},\ }\href
  {\doibase 10.1103/PhysRevA.75.033809} {\bibfield  {journal} {\bibinfo
  {journal} {Phys. Rev. A}\ }\textbf {\bibinfo {volume} {75}},\ \bibinfo
  {pages} {033809} (\bibinfo {year} {2007}{\natexlab{b}})}\BibitemShut
  {NoStop}%
\bibitem [{\citenamefont {Wen}\ \emph {et~al.}(2008)\citenamefont {Wen},
  \citenamefont {Du}, \citenamefont {Zhang}, \citenamefont {Xiao},\ and\
  \citenamefont {Rubin}}]{wen.PRA.77.033816.2008}%
  \BibitemOpen
  \bibfield  {author} {\bibinfo {author} {\bibfnamefont {J.}~\bibnamefont
  {Wen}}, \bibinfo {author} {\bibfnamefont {S.}~\bibnamefont {Du}}, \bibinfo
  {author} {\bibfnamefont {Y.}~\bibnamefont {Zhang}}, \bibinfo {author}
  {\bibfnamefont {M.}~\bibnamefont {Xiao}}, \ and\ \bibinfo {author}
  {\bibfnamefont {M.~H.}\ \bibnamefont {Rubin}},\ }\href {\doibase
  10.1103/PhysRevA.77.033816} {\bibfield  {journal} {\bibinfo  {journal} {Phys.
  Rev. A}\ }\textbf {\bibinfo {volume} {77}},\ \bibinfo {pages} {033816}
  (\bibinfo {year} {2008})}\BibitemShut {NoStop}%
\bibitem [{\citenamefont {Li}\ \emph {et~al.}(2024)\citenamefont {Li},
  \citenamefont {Wen}, \citenamefont {Cai}, \citenamefont {Ghamsari},
  \citenamefont {Li}, \citenamefont {Li}, \citenamefont {Zhang}, \citenamefont
  {Zhang},\ and\ \citenamefont {Xiao}}]{li2024direct}%
  \BibitemOpen
  \bibfield  {author} {\bibinfo {author} {\bibfnamefont {K.}~\bibnamefont
  {Li}}, \bibinfo {author} {\bibfnamefont {J.}~\bibnamefont {Wen}}, \bibinfo
  {author} {\bibfnamefont {Y.}~\bibnamefont {Cai}}, \bibinfo {author}
  {\bibfnamefont {S.~V.}\ \bibnamefont {Ghamsari}}, \bibinfo {author}
  {\bibfnamefont {C.}~\bibnamefont {Li}}, \bibinfo {author} {\bibfnamefont
  {F.}~\bibnamefont {Li}}, \bibinfo {author} {\bibfnamefont {Z.}~\bibnamefont
  {Zhang}}, \bibinfo {author} {\bibfnamefont {Y.}~\bibnamefont {Zhang}}, \ and\
  \bibinfo {author} {\bibfnamefont {M.}~\bibnamefont {Xiao}},\ }\href
  {https://www.science.org/doi/full/10.1126/sciadv.ado3199} {\bibfield
  {journal} {\bibinfo  {journal} {Sci. Adv.}\ }\textbf {\bibinfo {volume}
  {10}},\ \bibinfo {pages} {eado3199} (\bibinfo {year} {2024})}\BibitemShut
  {NoStop}%
\bibitem [{\citenamefont {Feng}\ \emph {et~al.}(2025)\citenamefont {Feng},
  \citenamefont {Zhuang}, \citenamefont {Liu}, \citenamefont {Liu},
  \citenamefont {Li},\ and\ \citenamefont {Zhang}}]{feng2025observation}%
  \BibitemOpen
  \bibfield  {author} {\bibinfo {author} {\bibfnamefont {Z.}~\bibnamefont
  {Feng}}, \bibinfo {author} {\bibfnamefont {R.}~\bibnamefont {Zhuang}},
  \bibinfo {author} {\bibfnamefont {S.}~\bibnamefont {Liu}}, \bibinfo {author}
  {\bibfnamefont {G.}~\bibnamefont {Liu}}, \bibinfo {author} {\bibfnamefont
  {K.}~\bibnamefont {Li}}, \ and\ \bibinfo {author} {\bibfnamefont
  {Y.}~\bibnamefont {Zhang}},\ }\href
  {https://advanced.onlinelibrary.wiley.com/doi/full/10.1002/advs.202501626}
  {\bibfield  {journal} {\bibinfo  {journal} {Adv. Sci.}\ }\textbf {\bibinfo
  {volume} {12}},\ \bibinfo {pages} {2501626} (\bibinfo {year}
  {2025})}\BibitemShut {NoStop}%
\bibitem [{\citenamefont {Li}\ \emph {et~al.}(2020)\citenamefont {Li},
  \citenamefont {Cai}, \citenamefont {Wu}, \citenamefont {Liu}, \citenamefont
  {Xiong}, \citenamefont {Li},\ and\ \citenamefont
  {Zhang}}]{kangkang.aqt.35.2020}%
  \BibitemOpen
  \bibfield  {author} {\bibinfo {author} {\bibfnamefont {K.}~\bibnamefont
  {Li}}, \bibinfo {author} {\bibfnamefont {Y.}~\bibnamefont {Cai}}, \bibinfo
  {author} {\bibfnamefont {J.}~\bibnamefont {Wu}}, \bibinfo {author}
  {\bibfnamefont {Y.}~\bibnamefont {Liu}}, \bibinfo {author} {\bibfnamefont
  {S.}~\bibnamefont {Xiong}}, \bibinfo {author} {\bibfnamefont
  {Y.}~\bibnamefont {Li}}, \ and\ \bibinfo {author} {\bibfnamefont
  {Y.}~\bibnamefont {Zhang}},\ }\href {\doibase
  https://doi.org/10.1002/qute.201900119} {\bibfield  {journal} {\bibinfo
  {journal} {Adv. Quantum Technol.}\ }\textbf {\bibinfo {volume} {3}},\
  \bibinfo {pages} {1900119} (\bibinfo {year} {2020})}\BibitemShut {NoStop}%
\bibitem [{\citenamefont {Ling}\ \emph {et~al.}(2025)\citenamefont {Ling},
  \citenamefont {Zhao}, \citenamefont {Zhang},\ and\ \citenamefont
  {Wu}}]{wu.pra.112.013706.2025}%
  \BibitemOpen
  \bibfield  {author} {\bibinfo {author} {\bibfnamefont {X.-Y.}\ \bibnamefont
  {Ling}}, \bibinfo {author} {\bibfnamefont {H.-M.}\ \bibnamefont {Zhao}},
  \bibinfo {author} {\bibfnamefont {X.-J.}\ \bibnamefont {Zhang}}, \ and\
  \bibinfo {author} {\bibfnamefont {J.-H.}\ \bibnamefont {Wu}},\ }\href
  {\doibase 10.1103/5c5q-1tk8} {\bibfield  {journal} {\bibinfo  {journal}
  {Phys. Rev. A}\ }\textbf {\bibinfo {volume} {112}},\ \bibinfo {pages}
  {013706} (\bibinfo {year} {2025})}\BibitemShut {NoStop}%
\bibitem [{\citenamefont {Wen}\ \emph {et~al.}(2010)\citenamefont {Wen},
  \citenamefont {Oh},\ and\ \citenamefont {Du}}]{Wen:10}%
  \BibitemOpen
  \bibfield  {author} {\bibinfo {author} {\bibfnamefont {J.}~\bibnamefont
  {Wen}}, \bibinfo {author} {\bibfnamefont {E.}~\bibnamefont {Oh}}, \ and\
  \bibinfo {author} {\bibfnamefont {S.}~\bibnamefont {Du}},\ }\href {\doibase
  10.1364/JOSAB.27.000A11} {\bibfield  {journal} {\bibinfo  {journal} {J. Opt.
  Soc. Am. B}\ }\textbf {\bibinfo {volume} {27}},\ \bibinfo {pages} {A11}
  (\bibinfo {year} {2010})}\BibitemShut {NoStop}%
\bibitem [{\citenamefont {Shalm}\ \emph {et~al.}(2012)\citenamefont {Shalm},
  \citenamefont {Hamel}, \citenamefont {Yan}, \citenamefont {Simon},\ and\
  \citenamefont {Jennewein}}]{shalm.np.9.1.2012}%
  \BibitemOpen
  \bibfield  {author} {\bibinfo {author} {\bibfnamefont {L.~K.}\ \bibnamefont
  {Shalm}}, \bibinfo {author} {\bibfnamefont {D.~R.}\ \bibnamefont {Hamel}},
  \bibinfo {author} {\bibfnamefont {Z.}~\bibnamefont {Yan}}, \bibinfo {author}
  {\bibfnamefont {C.}~\bibnamefont {Simon}}, \ and\ \bibinfo {author}
  {\bibfnamefont {T.}~\bibnamefont {Jennewein}},\ }\href
  {https://www.nature.com/articles/nphys2492} {\bibfield  {journal} {\bibinfo
  {journal} {Nat. Phys.}\ }\textbf {\bibinfo {volume} {9}},\ \bibinfo {pages}
  {19} (\bibinfo {year} {2012})}\BibitemShut {NoStop}%
\bibitem [{\citenamefont {Hubel}\ \emph {et~al.}(2010)\citenamefont {Hubel},
  \citenamefont {Hamel}, \citenamefont {Fedrizzi}, \citenamefont {Ramelow},
  \citenamefont {Resch},\ and\ \citenamefont {Jennewein}}]{hubel2010direct}%
  \BibitemOpen
  \bibfield  {author} {\bibinfo {author} {\bibfnamefont {H.}~\bibnamefont
  {Hubel}}, \bibinfo {author} {\bibfnamefont {D.~R.}\ \bibnamefont {Hamel}},
  \bibinfo {author} {\bibfnamefont {A.}~\bibnamefont {Fedrizzi}}, \bibinfo
  {author} {\bibfnamefont {S.}~\bibnamefont {Ramelow}}, \bibinfo {author}
  {\bibfnamefont {K.~J.}\ \bibnamefont {Resch}}, \ and\ \bibinfo {author}
  {\bibfnamefont {T.}~\bibnamefont {Jennewein}},\ }\href
  {https://www.nature.com/articles/nature09175} {\bibfield  {journal} {\bibinfo
   {journal} {Nature}\ }\textbf {\bibinfo {volume} {466}},\ \bibinfo {pages}
  {601} (\bibinfo {year} {2010})}\BibitemShut {NoStop}%
\bibitem [{\citenamefont {Zhang}\ \emph {et~al.}(2011)\citenamefont {Zhang},
  \citenamefont {Chen}, \citenamefont {Liu}, \citenamefont {Loy}, \citenamefont
  {Wong},\ and\ \citenamefont {Du}}]{du.prl.106.243602.2011}%
  \BibitemOpen
  \bibfield  {author} {\bibinfo {author} {\bibfnamefont {S.}~\bibnamefont
  {Zhang}}, \bibinfo {author} {\bibfnamefont {J.~F.}\ \bibnamefont {Chen}},
  \bibinfo {author} {\bibfnamefont {C.}~\bibnamefont {Liu}}, \bibinfo {author}
  {\bibfnamefont {M.~M.~T.}\ \bibnamefont {Loy}}, \bibinfo {author}
  {\bibfnamefont {G.~K.~L.}\ \bibnamefont {Wong}}, \ and\ \bibinfo {author}
  {\bibfnamefont {S.}~\bibnamefont {Du}},\ }\href {\doibase
  10.1103/PhysRevLett.106.243602} {\bibfield  {journal} {\bibinfo  {journal}
  {Phys. Rev. Lett.}\ }\textbf {\bibinfo {volume} {106}},\ \bibinfo {pages}
  {243602} (\bibinfo {year} {2011})}\BibitemShut {NoStop}%
\bibitem [{\citenamefont {Liao}\ \emph {et~al.}(2014)\citenamefont {Liao},
  \citenamefont {Yan}, \citenamefont {He}, \citenamefont {Du}, \citenamefont
  {Zhang},\ and\ \citenamefont {Zhu}}]{PhysRevLett.112.243602}%
  \BibitemOpen
  \bibfield  {author} {\bibinfo {author} {\bibfnamefont {K.}~\bibnamefont
  {Liao}}, \bibinfo {author} {\bibfnamefont {H.}~\bibnamefont {Yan}}, \bibinfo
  {author} {\bibfnamefont {J.}~\bibnamefont {He}}, \bibinfo {author}
  {\bibfnamefont {S.}~\bibnamefont {Du}}, \bibinfo {author} {\bibfnamefont
  {Z.-M.}\ \bibnamefont {Zhang}}, \ and\ \bibinfo {author} {\bibfnamefont
  {S.-L.}\ \bibnamefont {Zhu}},\ }\href {\doibase
  10.1103/PhysRevLett.112.243602} {\bibfield  {journal} {\bibinfo  {journal}
  {Phys. Rev. Lett.}\ }\textbf {\bibinfo {volume} {112}},\ \bibinfo {pages}
  {243602} (\bibinfo {year} {2014})}\BibitemShut {NoStop}%
\end{thebibliography}

%

\end{document}